\documentclass{article}
\usepackage{amssymb}

\newcommand{\be}{\begin{equation}}
\newcommand{\ee}{\end{equation}}
\newcommand{\bea}{\begin{eqnarray}}
\newcommand{\eea}{\end{eqnarray}}
\newcommand{\ben}{\begin{eqnarray*}}
\newcommand{\een}{\end{eqnarray*}}
\newcommand{\dst}{\displaystyle}

\newcommand{\pa}{\partial}

\newcommand{\ve}{\varepsilon}
\newcommand{\bns}{\beta_{\rm ns}}
\newcommand{\nonv}{({\bf n}_0\cdot{\bf n}_v)}
\newcommand{\nond}{({\bf n}_0\cdot{\bf n}_{\rm d})}
\newcommand{\ndnv}{({\bf n}_{\rm d}\cdot{\bf n}_v)}

\def\ssbi#1{{\stackrel{\scriptscriptstyle(#1)}{f}}_{\!\!{\rm SSB}}}
\def\ssbii#1{{\stackrel{\scriptscriptstyle(#1)}{f}}_{\!\!0}}
\def\nssd#1{\stackrel{\scriptscriptstyle(#1)}{f_{\rm ns}}}

\begin{document}

\title{
Data analysis of gravitational-wave signals
\\from spinning neutron stars.
I. The signal and its detection}

\author{Piotr Jaranowski$^{1,3}$
\and Andrzej Kr\'olak$^{1,4}$
\and Bernard F.\ Schutz$^{1,2}$
\\{\it $^1$Albert-Einstein-Institut,
       Max-Planck-Institut f\"ur Gravitationsphysik}\\
  {\it Schlaatzweg 1, 14473 Potsdam, Germany}
\\{\it $^2$Department of Physics and Astronomy,
       University of Wales College of Cardiff}\\
  {\it PO BOX 913, Cardiff, U.K.}
\\{\it $^3$Institute of Physics, Bia{\l}ystok University}\\
  {\it Lipowa 41, 15-424 Bia{\l}ystok, Poland}
\\{\it $^4$Institute of Mathematics,
       Polish Academy of Sciences}\\
  {\it \'{S}niadeckich 8, 00-950 Warsaw, Poland}}
\maketitle

\begin{abstract}

We present a theoretical background for the data analysis of the
gravitational-wave signals from spinning neutron stars for Earth-based laser
interferometric detectors.  We introduce a detailed model of the signal
including both the frequency and the amplitude modulations.  We include the
effects of the intrinsic frequency changes and the modulation of the frequency
at the detector due to the Earth motion.  We estimate the effects of the star's
proper motion and of relativistic corrections.  Moreover we consider a signal
consisting of two components corresponding to a frequency $f$ and twice that
frequency.  From the maximum likelihood principle we derive the detection
statistics for the signal and we calculate the probability density function of
the statistics.  We obtain the data analysis procedure to detect the signal and
to estimate its parameters.  We show that for optimal detection of the amplitude
modulated signal we need four linear filters instead of one linear filter needed
for a constant amplitude signal.  Searching for the doubled frequency signal
increases further the number of linear filters by a factor of two.  We indicate
how the fast Fourier transform algorithm and resampling methods commonly
proposed in the analysis of periodic signals can be used to calculate the
detection statistics for our signal.  We find that the probability density
function of the detection statistics is determined by one parameter:  the
optimal signal-to-noise ratio.  We study the signal-to-noise ratio by means of
the Monte Carlo method for all long-arm interferometers that are currently under
construction.  We show how our analysis can be extended to perform a joint
search for periodic signals by a network of detectors and we perform Monte Carlo
study of the signal-to-noise ratio for a network of detectors.

\end{abstract}

\section{Introduction}

Spinning neutron stars are one of the primary candidate sources of gravitational
waves for long-arm laser interferometric detectors (\cite{S96a}, see \cite{T87}
for a review).  Detectors with a sufficent sensitivity to see strong neutron
star sources anywhere in the Galaxy will be taking data within two or three
years \cite{GEO600,LIGO,VIRGO,TAMA300}.  A rotating body, perfectly symmetric
about its rotation axis does not emit gravitational waves.  If the spinning
neutron star is to emit gravitational waves over extended periods of time, it
must have some kind of long-lived asymmetry.  Several mechanisms have been given
for such an asymmetry to arise \cite{PPS76,BG96,ZS79,Z80}.  During the
crystalization period the crust of the neutron star may develop deviations from
axisymmetry that will be supported by anisotropic stresses in the solid crust
\cite{PPS76}.  The strong magnetic field present in the neutron star may not be
aligned with the rotation axis and consequently the distortion produced by the
magnetic pressure results in the neutron star being asymmetric \cite{BG96}.
Also the rotation axis may not coincide with a principal axis of the star's
moment of inertia tensor.  Then the star will precess and emit gravitational
waves \cite{ZS79,Z80}.  There are other mechanisms that can produce
gravitational waves from neutron stars.  Accretion of matter on a neutron star
can drive it into a nonaxisymmetric configuration and power steady radiation
with a considerable amplitude.  This mechanism has been pointed out by Wagoner
(\cite{W84}, see also \cite{S96b}).  It applies to a certain class of neutron
stars, including accreting stars in binary systems that have been spun up to the
first instability point of the Chandrasekhar-Friedman-Schutz (CFS) instability
\cite{C70,FS78}.  Recently Andersson \cite{And97} suggested a similar
instability in $r-$modes of rotating relativistic stars.  The effectiveness of
these instabilities depends on the viscocity of the star which in turn is
determined by the temperature of the star \cite{Lin97}.

This paper initiates a series of papers where theoretical problems of data
analysis of gra\-vi\-ta\-tio\-nal-wave signals from spinning neutron stars are
considered, independently of the mechanisms generating the waves.

The data analysis of monochromatic signals for interferometric antennae was
investigated by one of us \cite{S91}.  A search strategy for such signals was
proposed and the computing power required estimated.  The basic method to detect
periodic signals is to Fourier analyse the data, and an efficient computational
tool is the fast Fourier transform.  The main problem is that to do the search
one has to take into account the modulation of the signal due to the Earth's
motion relative to the solar system barycenter.  If the position of the source
on the sky is unknown this introduces two additional parameters in the signal
and this vastly increases the computational time to do the search.  It is clear
that the main limit on the sensitivity of such a search will be the available
computing power.  Variants of the proposed search strategy have been implemented
with test data from the prototype detectors where the search was carried out
only over a limited region of the parameter space \cite{L87,N93,J95}.

The problem of computational requirements has recently been reconsidered by
Brady et al.\ \cite{BCCS97}.  They realized that in the model of the signal the
effect of the intrinsic frequency modulation due to spin-down or spin-up of the
neutron star needs to be considered.  This increases the parameter space and
consequently the computational power required to search all the parameter space.
Assuming access to teraflops computing power it was shown that coherent
integration times will be limited to days for an all-sky search for young,
rapidly spinning stars and to weeks for more directed searches.  A simplified
model of the signal where modulation due to diurnal rotation of the Earth was
neglected has also been examined by one of us \cite{K97a} and the computational
requirements to do the search were estimated.

In this series of papers we consider a more general model of the signal than in
the work cited above.  We take into account not only the modulation of the phase
of the signal but also the amplitude modulation.  Moreover we consider a signal
consisting of two components corresponding to a frequency $f$ and twice that
frequency.  In general neither of the components is dominant.

In this work, which is Paper I of the series, we introduce the signal and we
derive an optimal data analysis procedure for its detection.  In Paper II we
examine the accuracy of estimation of the parameters of the signal that can be
achieved with the optimal analysis.  In Paper III we examine in detail the
characteristics of the detection statistics derived in Paper I and the
computational power required to calculate it.  In Paper IV we investigate
the least-squares method to estimate astrophysically interesting parameters 
of the signal from the estimators of the amplitudes derived in Paper I.

The plan of this paper is as follows.  In Section 2 we derive a general formula
for the response of a laser interferometer to our two component signal including
both the phase and the amplitude modulation.  In Section 3 from the maximum
likelihood principle we derive the data analysis procedure to detect the signal
introduced in Section 2 and to estimate its parameters.  We obtain the basic
probability density functions of the detection statistics.  We show that
probability of detection is determined by one parameter:  the optimal
signal-to-noise ratio.  We study this quantity by means of the Monte Carlo
simulations for all the interferometric detectors that are currently under
construction.  We conclude Section 3 by showing how one can take advantage of
the speed of the FFT algorithm to evaluate efficiently our detection statistics.
This involves application of the resampling techniques proposed earlier for the
case of a simpler signal model \cite{S91,BCCS97}.  In Section 4 we show that our
analysis can easily be extended to networks of detectors and we perform a Monte
Carlo study of the signal-to-noise ratio for the networks.  In Appendix A we
discuss the model of the phase of the gravitational-wave signal and in
particular we estimate the effect of the proper motion of the neutron star and 
of
relativistic corrections.  In Appendix B we give the general analytic formula
for the optimal signal-to-noise ratio.  In Appendix C we present an additional
study by means of the Monte Carlo simulations of the signal-to-noise ratios for the
individual components of the spinning neutron star signal.

\section{Noise-free response of the interferometric detector}

\subsection{Beam-pattern functions}

The response of a laser interferometric detector to a weak plane gravitational 
wave in the long wavelength approximation (i.e.\ when the size of the detector 
is much smaller than the reduced wavelength $\lambda/(2\pi)$ of the wave) is 
well known (see, e.g., \cite{ST87} and Section~IIA of \cite{JK94} 
and references therein). The dimensionless detector's 
response function $h$ is defined as the difference between the wave induced 
relative length changes of the two interferometer arms and can be computed from 
the formula (cf.\ Eq.\ (5) of \cite{JK94})
\be
\label{bp1}
h(t)=\frac{1}{2}{\bf n}_1\cdot\left[\widetilde{H}(t){\bf n}_1\right]
-\frac{1}{2}{\bf n}_2\cdot\left[\widetilde{H}(t){\bf n}_2\right],
\ee
where ${\bf n}_1$ and ${\bf n}_2$ denote the unit vectors parallel to the arm
number 1 and 2, respectively (the order of arms if defined such that the vector
${\bf n}_1\times{\bf n}_2$ points {\em outwards} from the surface of the Earth),
$\widetilde{H}$ is the 3-dimensional matrix of the spatial metric perturbation
produced by the wave in the proper reference frame of the detector, and a dot 
stands for the standard scalar product in the 3-dimensional Cartesian space.  
The matrix $\widetilde{H}$ is given by
\be
\label{bp2}
\widetilde{H}(t)=M(t)H(t)M(t)^T,
\ee
where $M$ is the 3-dimensional orthogonal matrix of transformation from the
wave Cartesian coordinates $(x_w,y_w,z_w)$ to the Cartesian coordinates
$(x_{\rm d},y_{\rm d},z_{\rm d})$ in the detector's proper reference frame (the 
definition of 
these coordinates is given below), $T$ denotes matrix 
transposition.  In the wave coordinate system the 
gravitational wave travels in the $+z_w$ direction. 
In this frame the matrix $H$ has the form
\be
\label{bp3}
H(t)=\left(\begin{array}{ccc}
h_+(t) & h_{\times}(t) & 0 \\
h_{\times}(t) & -h_+(t) & 0 \\
0 & 0 & 0
\end{array}\right),
\ee
where the functions $h_+$ and $h_\times$ describe two independent wave's
polarizations.  Collecting Eqs.\ (\ref{bp1})--(\ref{bp3}) together one can see
that the response function $h$ is a linear combination of the functions $h_+$ 
and $h_\times$:
\be
\label{bp4}
h(t)=F_+(t)h_+(t)+F_{\times}(t)h_{\times}(t),
\ee
where $F_+$ and $F_\times$ are called the {\em beam-pattern} functions.

Because of the diurnal motion of the Earth the beam-patterns $F_+$ and
$F_\times$ are periodic functions of time with a period equal to one sidereal
day.  We want now to extract explicitly this time dependence as well as to
express $F_+$ and $F_\times$ as functions of right ascension $\alpha$ and
declination $\delta$ of the gravitational-wave source and polarization angle
$\psi$ (the angles $\alpha$, $\delta$, and $\psi$ determine the orientation of 
the wave reference frame with respect to the celestial sphere reference frame 
defined below).  Our treatment partially follows that of Section~5 of 
\cite{BG96}.  We represent the matrix $M$ of Eq.\ (\ref{bp2}) as
\be
M = M_3\,M_2\,M_1^T,
\ee
where $M_1$ is the matrix of transformation from wave to celestial sphere frame
coordinates, $M_2$ is the matrix of transformation from celestial coordinates to
cardinal coordinates and $M_3$ is the matrix of transformation from cardinal
coordinates to detector proper reference frame coordinates.  In celestial sphere
coordinates the $z$ axis coincides with the Earth's rotation axis and points
toward the North pole, the $x$ and $y$ axes lie in the Earth's equatorial plane,
and the $x$ axis points toward the vernal point.  In cardinal coordinates the
$(x,y)$ plane is tangent to the surface of the Earth at detector's location with
$x$ axis in the North-South direction and $y$ axis in the West-East direction,
the $z$ cardinal axis is along the Earth's radius pointing toward zenith.  In
detector coordinates the $z$ axis coincides with the $z$ axis of cardinal
coordinates and the $x$ axis is along the first interferometer arm (then the $y$
axis is along the second arm if the arms are at a right angle).  Under the above
conventions the matrices $M_1$, $M_2$, and $M_3$ are as follows (matrices $M_1$
and $M_2$ given below coincide with matrices $A$ and $B$ from Ref.\ \cite{BG96},
cf.\ Eqs.\ (52) and (60) of \cite{BG96})
\bea
\label{m1}
M_1&=&\left(\begin{array}{ccc}
\sin\alpha\cos\psi-\cos\alpha\sin\delta\sin\psi&
-\cos\alpha\cos\psi-\sin\alpha\sin\delta\sin\psi&
\cos\delta\sin\psi\\ 
-\sin\alpha\sin\psi-\cos\alpha\sin\delta\cos\psi&
\cos\alpha\sin\psi-\sin\alpha\sin\delta\cos\psi&
\cos\delta\cos\psi\\ 
-\cos\alpha\cos\delta&
-\sin\alpha\cos\delta&
-\sin\delta
\end{array}\right),\quad\\
\label{m2}
M_2&=&\left(\begin{array}{ccc}
\sin\lambda\cos(\phi_r+\Omega_r t)&
\sin\lambda\sin(\phi_r+\Omega_r t)&
-\cos\lambda\\
-\sin(\phi_r+\Omega_r t)&
\cos(\phi_r+\Omega_r t)&0\\
\cos\lambda\cos(\phi_r+\Omega_r t)&
\cos\lambda\sin(\phi_r+\Omega_r t)&
\sin\lambda
\end{array}\right),\\
\label{m3}
M_3 &=& \left(\begin{array}{ccc}
-\sin\left(\gamma+\zeta/2\right) &
\cos\left(\gamma+\zeta/2\right) & 0 \\
-\cos\left(\gamma+\zeta/2\right) &
-\sin\left(\gamma+\zeta/2\right) & 0 \\
0 & 0 & 1
\end{array}\right).
\eea
In Eq.\ (\ref{m2}) $\lambda$ is the latitude of the detector's site, $\Omega_r$
is the rotational angular velocity of the Earth, and $\phi_r$ is a deterministic
phase which defines the position of the Earth in its diurnal motion at $t=0$
(the sum $\phi_r+\Omega_r t$ coincides with the local sidereal time of the
detector's site, i.e.\ with the angle between the local meridian and the vernal
point).  In Eq.\ (\ref{m3}) $\gamma$ determines the orientation of the
detector's arms with respect to local geographical directions:  $\gamma$ is
measured counter-clockwise from East to the bisector of the interferometer arms,
and $\zeta$ is the angle between the interferometer arms.  The vectors ${\bf
n}_1$ and ${\bf n}_2$ from Eq.\ (\ref{bp1}) in the detector's reference frame
have coordinates
\be
\label{n1n2}
{\bf n}_1=\left(1,0,0\right),\quad{\bf n}_2
=\left(\cos\zeta,\sin\zeta,0\right).
\ee
The values of the angles $\lambda$, $\gamma$, $\zeta$, and the longitudes $L$ 
(measured positively westwards) for different detectors can be found in Table 1 
\cite{A96}.

\vspace{2ex}\begin{table}[h]
\begin{center}
\begin{tabular}{|c|c|c|c|c|}\hline
detector & $\lambda$ (degrees) & $L$ (degrees)
& $\gamma$ (degrees) & $\zeta$ (degrees)\\ 
\hline
GEO600          & 52.25 &   $-$9.81 &  68.775 & 94.33 \\ \hline
LIGO Hanford    & 46.45 &  119.41 & 171.8 &  90   \\ \hline
LIGO Livingston & 30.56 &   90.77 & 243.0 &  90   \\ \hline
VIRGO           & 43.63 &  $-$10.5  & 116.5 &  90   \\ \hline
TAMA300         & 35.68 & $-$139.54 & 225.0 &  90   \\ \hline
\end{tabular}
\end{center}
\caption{Positions and orientations of detectors.}
\end{table}

To find the explicit formula for $F_+$ and $F_\times$ we have to combine Eqs.\ 
(\ref{bp1})--(\ref{n1n2}). After extensive algebraic manipulations we arrive at
the expressions: 
\bea
\label{bp5a}
F_+(t) &=& \sin\zeta\left[a(t)\cos2\psi+b(t)\sin2\psi\right],\\
\label{bp5b}
F_\times(t) &=& \sin\zeta\left[b(t)\cos2\psi-a(t)\sin2\psi\right],
\eea
where
\bea
\label{adef}
a(t) &=&
\frac{1}{16}\sin2\gamma(3 - \cos2\lambda)(3 - \cos2\delta)
\cos[2(\alpha-\phi_r-\Omega_r t)]
\nonumber\\&&
-\frac{1}{4}\cos2\gamma\sin\lambda(3 - \cos2\delta)
\sin[2(\alpha-\phi_r-\Omega_r t)]
\nonumber\\&&
+\frac{1}{4}\sin2\gamma\sin2\lambda\sin2\delta
\cos[\alpha-\phi_r-\Omega_r t]
\nonumber\\&&
-\frac{1}{2}\cos2\gamma\cos\lambda\sin2\delta
\sin[\alpha-\phi_r-\Omega_r t]
\nonumber \\&&
+\frac{3}{4}\sin2\gamma\cos^2\lambda\cos^2\delta,
\\
\label{bdef}
b(t) &=&
\cos2\gamma\sin\lambda\sin\delta\cos[2(\alpha-\phi_r-\Omega_r t)]
\nonumber \\&&
+\frac{1}{4}\sin2\gamma(3 - \cos2\lambda)\sin\delta
\sin[2(\alpha-\phi_r-\Omega_r t)]
\nonumber \\&&
+\cos2\gamma\cos\lambda\cos\delta\cos[\alpha-\phi_r-\Omega_r t]
\nonumber \\&&
+\frac{1}{2}\sin2\gamma\sin2\lambda\cos\delta
\sin[\alpha-\phi_r-\Omega_r t].
\eea
By means of Eqs.\ (\ref{bp5a})--(\ref{bdef}) the beam-pattern functions
can be computed directly for any instant of time.

Equivalent explicit formulae for the beam-pattern functions $F_+$ and $F_\times$
(for the case $\zeta=\pi/2$) can be found in Ref.\ \cite{JD94} where different
angles describing the position of the gravitational-wave source in the sky and
the orientation of the detector on the Earth are used. Also for the case 
$\zeta=\pi/2$ the functions $a$ and $b$ can be found in Ref.\ \cite{N93}, where 
still another set of angles is used \cite{N93a}.

\subsection{The phase of the gravitational-wave signal}

In Appendix A we derive the time dependence of the phase of the
gravitational-wave signal observed at the detector's location.  We consider the
significance of the corrections due to the motion of both the detector and the
neutron star with respect to the the solar system barycenter (SSB) reference
frame as well as the importance of relativistic corrections.  On the basis of
the discussion presented in Appendix A we adopt the following model of the phase
of the gravitational-wave signal:
\be
\label{pha3}
\Psi(t) = \Phi_0
+ 2\pi\sum_{k=0}^{s}{\ssbii k}\frac{t^{k+1}}{(k+1)!}
+ \frac{2\pi}{c}{\bf n}_0\cdot{\bf r}_{\rm d}(t) 
\sum_{k=0}^{s}{\ssbii k}\frac{t^k}{k!},
\ee
where ${\ssbii k}$ is the $k$th time derivative of the instantaneous frequency
evaluated at $t = 0$ at the SSB, ${\bf n}_0$ is the constant unit vector in the
direction of the star in the SSB reference frame and ${\bf r}_{\rm d}$ is the
position vector of the detector in that frame.

The signal analysis presented in the remaining part of the paper does not depend
on the number $s$ of the spindown parameters and therefore we keep $s$
unspecified.

We associate a coordinate system with the SSB reference frame.  The $x$ axis of
the system is parallel to the $x$ axis of the celestial sphere coordinate
system, the $z$ axis is perpendicular to the ecliptic and coincides with the
orbital angular momentum vector of the Earth.  In that system the unit vector
${\bf n}_0$ pointing towards the star has the components
\be
\label{pha4}
{\bf n}_0=\left(\begin{array}{ccc}
1 &    0     & 0       \\
0 &  \cos\ve & \sin\ve \\
0 & -\sin\ve & \cos\ve
\end{array}\right)
\left(\begin{array}{c}
\cos\alpha\cos\delta\\ \sin\alpha\cos\delta\\ \sin\delta
\end{array}\right),
\ee
where $\ve$ is the angle between ecliptic and the Earth's equator. The position 
vector ${\bf r}_{\rm d}$ of the detector has in this coordinate system the 
components
\be
\label{pha5}
{\bf r}_{\rm d}=R_{ES}\left(\begin{array}{c}
\cos\left(\phi_o+\Omega_o t\right)\\ \sin\left(\phi_o+\Omega_o t\right) \\0
\end{array}\right)+R_E\left(\begin{array}{ccc}
1 &    0     & 0       \\
0 &  \cos\ve & \sin\ve \\
0 & -\sin\ve & \cos\ve
\end{array}\right)
\left(\begin{array}{c}
\cos\lambda\cos\left(\phi_r+\Omega_r t\right)\\
\cos\lambda\sin\left(\phi_r+\Omega_r t\right)\\ \sin\lambda
\end{array}\right),
\ee
where $R_{ES} =$ 1 AU is the mean distance from the Earth's center to the SSB,
$R_E$ is the mean radius of the Earth, $\Omega_o$ is the mean orbital angular
velocity of the Earth and $\phi_o$ is a deterministic phase which defines the
position of the Earth in its orbital motion at $t=0$. We recall that we neglect 
the eccentricity of the Earth's orbit and the motion of the Earth around the 
Earth-Moon barycenter.

Substituting Eqs.\ (\ref{pha4}) and (\ref{pha5}) into Eq.\ (\ref{pha3}) one 
gets
\bea
\label{pha6a}
\Psi(t)&=&\Phi_0+\Phi(t),\\
\label{pha6b}
\Phi(t)&=&2\pi\sum_{k=0}^{s}{\ssbii k}\frac{t^{k+1}}{(k+1)!}
\nonumber\\&&
+\frac{2\pi}{c}
\left\{R_{ES}\left[\cos\alpha\cos\delta\cos\left(\phi_o+\Omega_o t\right)
+\left(\cos\ve\sin\alpha\cos\delta+\sin\ve\sin\delta\right)
\sin\left(\phi_o+\Omega_o t\right)\right]
\right.\nonumber\\&&\left.
+R_{E}\left[\sin\lambda\sin\delta
+\cos\lambda\cos\delta\cos\left(\alpha-\phi_r-\Omega_r t\right)\right]\right\}
\sum_{k=0}^{s}{\ssbii k}\frac{t^k}{k!}.\qquad
\eea

\subsection{Wave polarization functions}

We use the following two-component model of the gravitational-wave signal:
\be
\label{s1}
h(t) = h_1(t) + h_2(t),
\ee
where
\bea
\label{s2}
&h_1(t) = F_+(t) h_{1+}(t) + F_\times(t) h_{1\times}(t),\quad
h_2(t) = F_+(t) h_{2+}(t) + F_\times(t) h_{2\times}(t),&\\
\label{s3}
&h_{1+}(t) = \frac{1}{8}h_o\sin2\theta\sin2\iota\cos\Psi(t),\quad
h_{2+}(t) = \frac{1}{2}h_o\sin^2\theta(1+\cos^2\iota)\cos2\Psi(t),&\\
\label{s4}
&h_{1\times}(t) = \frac{1}{4}h_o\sin2\theta\sin \iota\sin\Psi(t),\quad
h_{2\times}(t) = h_o\sin^2\theta\cos \iota\sin2\Psi(t).&
\eea
The beam-pattern functions $F_+$, $F_\times$ are given by Eqs.\ 
(\ref{bp5a})--(\ref{bdef}) and the phase $\Psi$ is given by Eqs.\ (\ref{pha6a}) 
and (\ref{pha6b}).

The model of the signal defined by Eqs.\ (\ref{s1})--(\ref{s4}) 
represents the quadrupole gravitational wave that is emitted by
a freely precessing axisymmetric star. The angle $\theta$,
called the wobble angle, is the angle between the total angular momentum vector
of the star and the star's axis of symmetry and $\iota$ is the angle between the
total angular momentum vector of the star and the direction from the star to the
Earth.  The amplitude $h_o$ is given by
\be
\label{ho}
h_o = \frac{16\pi^2G}{c^4}\frac{\epsilon I f^2}{r},
\ee
where $f$ is the sum of the frequency of rotation of the star and the frequency
of precession, $I$ is the moment of inertia with respect to the rotation axis,
$\epsilon$ is the poloidal ellipticity of the star and $r$ is the distance to
the star.  For small wobble angle the signal $h_1$ is dominant.  Details of the
model can be found in \cite{ZS79}.  When $\theta = \pi/2$ the $h_1$ component
vanishes.  For this special case the $h_2$ component is the quadrupole wave from
a triaxial ellipsoid rotating about a principal axis with frequency $f$.  In
this case the amplitude $h_o$ is again given by Eq.\ (\ref{ho}) except that
$\epsilon$ is now the ellipticity of the star defined by
\be
\epsilon = \frac{I_1 - I_2}{I},
\ee
where  $I_1$ and $I_2$ are the moments of inertia of the star with respect to 
the principal axes orthogonal to the rotation axis. This model was considered in 
\cite{BCCS97}.

Replacing the physical constants in Eq.\ (\ref{ho}) by their numerical values 
results in
\be
\label{hon}
h_o = 4.23\times10^{-25}d_o\left(\frac{f}{\rm 100~Hz}\right)^2,
\ee 
where
\be
\label{do}
d_o:=\left(\frac{\epsilon}{10^{-5}}\right)
\left(\frac{I}{10^{45}~\mbox{g cm}^2}\right)
\left(\frac{\mbox{1~kpc}}{r}\right).
\ee 

By means of Eqs.\ (\ref{bp5a}) and (\ref{bp5b}) the signal described by
Eqs.\ (\ref{s1})--(\ref{s4}) can be written in the form
\be
\label{sig}
h(t) = \sum^{4}_{i=1} A_{1i}\,h_{1i}(t)+\sum^{4}_{i=1} A_{2i}\,h_{2i}(t), 
\ee
where the eight amplitudes $A_{1i}$ and $A_{2i}$ are given by
\bea
A_{11}&=&h_o\sin\zeta\sin2\theta\left[
\frac{1}{8}\sin2\iota\cos2\psi\cos\Phi_0
-\frac{1}{4}\sin\iota\sin2\psi\sin\Phi_0
\right],
\\
A_{12}&=&h_o\sin\zeta\sin2\theta\left[
\frac{1}{4}\sin\iota\cos2\psi\sin\Phi_0
+\frac{1}{8}\sin2\iota\sin2\psi\cos\Phi_0
\right],
\\
A_{13}&=&h_o\sin\zeta\sin2\theta\left[
-\frac{1}{8}\sin2\iota\cos2\psi\sin\Phi_0
-\frac{1}{4}\sin\iota\sin2\psi\cos\Phi_0
\right],
\\
A_{14}&=&h_o\sin\zeta\sin2\theta\left[
\frac{1}{4}\sin\iota\cos2\psi\cos\Phi_0
-\frac{1}{8}\sin2\iota\sin2\psi\sin\Phi_0
\right],
\\
A_{21}&=&h_o\sin\zeta\sin^2\theta\left[
\frac{1}{2}(1 + \cos^2\iota)\cos2\psi\cos2\Phi_0
-\cos\iota\sin2\psi\sin2\Phi_0
\right],
\\
A_{22}&=&h_o\sin\zeta\sin^2\theta\left[
\frac{1}{2}(1 + \cos^2\iota)\sin2\psi\cos2\Phi_0
+\cos\iota\cos2\psi\sin2\Phi_0
\right],
\\
A_{23}&=&h_o\sin\zeta\sin^2\theta\left[
-\frac{1}{2}(1 + \cos^2\iota)\cos2\psi\sin2\Phi_0
-\cos \iota\sin2\psi\cos2\Phi_0
\right],
\\ 
A_{24}&=&h_o\sin\zeta\sin^2\theta\left[
-\frac{1}{2}(1 + \cos^2\iota)\sin2\psi\sin2\Phi_0
+\cos\iota\cos2\psi\cos2\Phi_0
\right].
\eea
The amplitudes $A_{1i}$ and $A_{2i}$ depend on the parameters $h_o, \theta,
\psi$, $\iota$, and $\Phi_0$.  They also depend on the angle $\zeta$.  The time
dependent functions $h_{li}$ have the form
\be
\begin{array}{rclrcl}
h_{l1} &=& a(t) \cos l\Phi(t), & h_{l2} &=& b(t) \cos l\Phi(t),\\
h_{l3} &=& a(t) \sin l\Phi(t), & h_{l4} &=& b(t) \sin l\Phi(t),
\end{array}\qquad l=1,2,
\ee
where the functions $a$ and $b$ are given by Eqs.\ (\ref{adef}) and
(\ref{bdef}), respectively, and $\Phi$ is the phase given by Eq.\ (\ref{pha6b}).
The modulation amplitudes $a$ and $b$ depend on the right ascension $\alpha$ and
the declination $\delta$ of the source (they also depend on the angles $\lambda$
and $\gamma$).  The phase $\Phi$ depends on the frequency $f_0$, $s$ spindown
parameters ${\ssbii k}$ $(k=1,\ldots,s)$, and on the angles $\alpha$,
$\delta$.  We call parameters $f_0$, ${\ssbii k}$, $\alpha$, $\delta$
the {\em phase parameters}.  Moreover the phase $\Phi$ depends on the
latitude $\lambda$ of the detector.  The whole signal $h$ depends on $8 + s$
unknown parameters:  $h_o, \theta, \psi, \iota, \Phi_0, \alpha, \delta, f_0,
{\ssbii k}$.

It is useful to consider the frequency domain characteristics of our
gravitational-wave signal.  The signal consists of two components with carrier
frequencies $f_0$ and $2f_0$ that are both amplitude and phase modulated.  The
amplitude modulation, determined by functions $a$ and $b$, splits each of the
two components into five lines corresponding to frequencies $f_0-2f_r$,
$f_0-f_r$, $f_0$, $f_0+f_r$, $f_0+2f_r$, where $f_r$ is the frequency of
rotation of Earth ($f_r\simeq$10$^{-5}$ Hz) and the same for frequency $2 f_0$.
The frequency modulation broadens the lines.  For the extreme case of the
gravitational-wave frequency of $10^3$ Hz, the spindown age $\tau = 40$ years,
and the observation time $T_o = 120$ days the maximum frequency shifts due to
the neutron star spindown, Earth orbital motion and Earth diurnal motion are
respectively $\sim$8 Hz, $\sim$0.1 Hz, and $\sim$10$^{-3}$ Hz.  As an example in
Figure 1 we have plotted the power spectrum of the noise-free response of a
detector located near Hannover to the gravitational wave from the Crab pulsar.
We took only the component $h_2$ with twice the rotational frequency.  We have
generated a 24-day long signal.

\section{Optimal filtering for the amplitude modulated signal}

\subsection{Maximum likelihood detection}

The signal given by Eq.\ (\ref{sig}) will be buried in the noise of a detector.
Thus we are faced with the problem of detecting the signal and estimating its
parameters.  A standard method is the method of {\em maximum likelihood
detection} which consists of maximizing the likelihood function $\Lambda$ with
respect to the parameters of the signal.  If the maximum of $\Lambda$ exceeds a
certain threshold calculated from the false alarm probability that we can afford
we say that the signal is detected.  The values of the parameters that maximize
$\Lambda$ are said to be the {\em maximum likelihood (ML) estimators} of the
parameters of the signal.  The magnitude of the maximum of $\Lambda$ determines
the probability of detection of the signal.

We assume that the noise $n$ in the detector is an additive, stationary, 
Gaussian, and zero-mean continuous random process. Then the data $x$ (if the 
signal $h$ is present) can be written as
\be
x(t) = n(t) + h(t).
\ee
The log likelihood function has the 
form
\be
\log\Lambda = (x|h) - \frac{1}{2}(h|h),
\ee
where the scalar product $(\,\cdot\,|\,\cdot\,)$ is defined by 
\be
\label{SP}
(x|y):=
4\Re\int^{\infty}_{0}\frac{\tilde{x}(f)\tilde{y}^{*}(f)}{S_h(f)}df,
\ee
where $\tilde{}$ denotes the Fourier transform, $*$ is complex conjugation, and 
$S_h$ is the {\em one-sided} spectral density of the detector's noise.

The gravitational-wave signal given by Eq.\ (\ref{sig}) consists of two narrowband
components around the frequencies $f_0$ and $2 f_0$ and therefore
to a very good accuracy the likelihood ratio is given by
\be
\label{LF}
\log\Lambda
\cong (x|h_1) - \frac{1}{2}(h_1|h_1) + (x|h_2) - \frac{1}{2}(h_2|h_2).
\ee
This suggests that we consider the two components of the response function
(\ref{sig}) as two independent signals.  Let us take the first component $h_1$
of the signal.  We can assume that over the bandwidth of the signal $S_h(f)$ is
nearly constant and equal to $S_h(f_0)$ where $f_0$ is the frequency of the
signal $h_1$ at $t=0$.  Thus in our case the above scalar product can be
approximated by
\be
(x|h_1) \cong \frac{2}{S_h(f_0)}\int^{T_o/2}_{-T_o/2}x(t)h_1(t)\,dt,
\ee
where $T_o$ is the observation time and where the observation interval is 
$\left[-T_o/2,T_o/2\right]$. It is useful to introduce the following 
scalar product 
\be
(x||y) := \frac{2}{T_o}\int^{T_o/2}_{-T_o/2}x(t)y(t)\,dt.
\ee
As long as the detector's noise is stationary over the observation period, this 
is a good scalar product. In realistic observations, the detector's noise will 
vary slowly during the observation period. We do not treat this important issue 
in this paper.

The log likelihood function for this signal is approximately given by
\be
\log\Lambda_1 \cong \frac{T_o}{S_h(f_0)}
\left[(x||h_1) - \frac{1}{2}(h_1||h_1)\right].
\ee
The maximum likelihood estimators can be found by maximizing the
following {\em normalized log likelihood function}
\be
\label{nlf}
\log\Lambda'_{1} = (x||h_1) - \frac{1}{2}(h_1||h_1).
\ee
The normalized log likelihood function does not involve 
explicitly the spectral density of the noise in the detector.

The signal $h_1$ depends linearly on four amplitudes $A_{1i}$.
The amplitudes depend on the five unknown parameters $h_o$, $\theta$, 
$\psi$, $\iota$, and $\Phi_0$ and are independent.
The likelihood equations for the amplitudes $A_{1i}$ are given by
\be
\frac{\partial\ln\Lambda'_1}{\partial A_{1i}} = 0,\quad i=1,\ldots,4.
\ee
One easily finds that in our case the above set of equations is equivalent
to the following set of linear algebraic equations
\be
\sum_{j=1}^4{\mathcal M}_{ij} A_{1j} = (x||h_{1i}),\quad i=1,\ldots,4,
\ee
where the components ${\mathcal M}_{ij}$ of the $4 \times 4$ matrix
${\bf \mathcal{M}}$ are given by
\be
\label{MLE}
{\mathcal M}_{ij} := (h_{1i}||h_{1j}).
\ee
Since over a typical observation time $T_o$ the phase $\Phi$ will
have very many oscillations, then to a very good accuracy we have
\be
(h_{11}||h_{13}) \cong 0,\quad
(h_{11}||h_{14}) \cong 0,\quad
(h_{12}||h_{13}) \cong 0,\quad
(h_{12}||h_{14}) \cong 0,
\ee
and also
\be\begin{array}{ccccc}
(h_{11}||h_{11}) &\cong& (h_{13}||h_{13}) &\cong& \frac{1}{2} A,\\
(h_{12}||h_{12}) &\cong& (h_{14}||h_{14}) &\cong& \frac{1}{2} B,\\ 
(h_{11}||h_{12}) &\cong& (h_{13}||h_{14}) &\cong& \frac{1}{2} C,
\end{array}\ee
where $A := (a||a)$, $B := (b||b)$, $C := (a||b)$.
With these approximations the matrix ${\bf {\mathcal M}}$ is given by
\bea
{\mathcal M} =
\left(\begin{array}{cc}
{\mathcal C} & {\mathcal O} \\
{\mathcal O} & {\mathcal C}
\end{array}\right),
\label{M}
\eea 
where ${\mathcal O}$ is a zero $2$ by $2$ matrix 
and ${\mathcal C}$ equals
\bea
{\mathcal C} = \frac{1}{2}\left( \begin{array}{cc}
A & C\\C & B\end{array} \right).
\label{C}      
\eea
Thus ${\bf {\mathcal M}}$ splits into two identical $2 \times 2$ matrices.
Assuming that $a \neq b$, $A \neq 0$ and $B \neq 0$ 
the explicit expressions for maximum likelihood estimators 
$\widehat{A}_{1i}$ of the amplitudes $A_{1i}$ are readily 
obtained and they are given by
\be
\label{amle}
\begin{array}{rcl}
\widehat{A}_{11} &=& {\dst 2\frac{B(x||h_{11}) - C(x||h_{12})}{D},}\\
\widehat{A}_{12} &=& {\dst 2\frac{A(x||h_{12}) - C(x||h_{11})}{D},}\\
\widehat{A}_{13} &=& {\dst 2\frac{B(x||h_{13}) - C(x||h_{14})}{D},}\\
\widehat{A}_{14} &=& {\dst 2\frac{A(x||h_{14}) - C(x||h_{13})}{D},}
\end{array}
\ee
where $D$ is defined by
\be
D = A B - C^2.
\ee
The second partial derivatives of the log likelihood function w.r.t.\ $A_{1i}$ 
are given by
\be
\frac{\partial^2\ln\Lambda'_1}
{\partial A_{1i}\partial A_{1j}} = -{\mathcal M}_{ij}.
\ee
Since $a \neq b$ it follows from Schwarz inequality that $D>0$. Thus as $A > 0$ 
and $B > 0$ the matrix ${\mathcal M}$ is positive-definite.
Therefore the extrema of the log likelihood function w.r.t.\  $A_{1i}$ are the 
local maxima.  The above ML estimators of the amplitudes $A_{1i}$ are 
substituted for the amplitudes $A_{1i}$ in the likelihood function (\ref{nlf}) 
giving the reduced normalized likelihood function $\Lambda''_1 = \exp({\mathcal 
F}_1)$ where ${\mathcal F}_1$ is given by
\bea
{\mathcal F}_1 &=& 
\frac{B (x||h_{11})^2 + A (x||h_{12})^2 - 2 C (x||h_{11})(x||h_{12})}{D}
\nonumber\\&&
+\frac{B (x||h_{13})^2 + A (x||h_{14})^2 - 2 C (x||h_{13})(x||h_{14})}{D}.
\eea
Thus to obtain the maximum likelihood estimators of the parameters of the signal
one first finds the maximum of the functional ${\cal F}_1$ with respect to the
frequency, the spindown parameters and the angles $\alpha$ and $\delta$ and then
one calculates the estimators of the amplitudes $A_{1i}$ from the analytic
formulae (\ref{amle}) with the correlations $(x||h_{1i})$ evaluated at the
values of the parameters obtained by the maximization of the functional
${\mathcal F}_1$.  Thus we see that filtering for the gravitational-wave signal
from a neutron star requires {\em four linear} filters.  Efficient numerical
methods to calculate the statistics ${\mathcal F}_1$ are discussed in
Section 3.4.

Exactly the same procedure applies to the second component of the 
signal. The formulae for the estimators of the amplitudes $A_{2i}$ and the
normalized reduced statistics ${\mathcal F}_2$ are obtained from the above
formulae by replacing $h_{1i}$ by $h_{2i}$.

To consider the optimal detection of the whole two-component signal we need to
remember that the eight amplitudes $A_{li}$ are not independent.  They depend on
five parameters:  $h_o$, $\theta$, $\psi$, $\iota$, and $\Phi_0$.  To find the
maximum likelihood estimators of the independent five parameters we would have
to maximize the total likelihood function (given by Eq.\ (\ref{LF})) with
respect to these parameters.  This however leads to an intractable set of
nonlinear algebraic equations which would have to be solved numerically, thereby
increasing the computational cost of the search for the signal.  Instead we
propose the following procedure.

We form the statistics
\be
{\mathcal F}
=\frac{T_o}{S_h(f_0)}{\mathcal F}_1
+\frac{T_o}{S_h(2f_0)}{\mathcal F}_2.
\label{STAT}
\ee
This is just the reduced likelihood function assuming that the eight 
amplitudes are independent.  
We first maximize the functional ${\mathcal F}$ with respect to
the frequency, spindown parameters and angles $\alpha$ and $\delta$ and we
calculate the eight amplitudes from the analytic formulae.  We then find the
estimators of the five independent parameters from the estimators of the
amplitudes by least-squares method.  We use the inverse of Fisher matrix for the
covariance matrix in the least-squares method.  We shall consider this problem
in Paper IV.

To announce the detection of the signal the functional ${\mathcal F}$ must
exceed a certain threshold calculated on the basis of the false alarm
probability that one can afford.  Once ${\mathcal F}$ is above the threshold its
magnitude determines the probability of detection of the signal.  Consequently
we need to determine the probability density function of ${\mathcal F}$ both
when the signal is absent and present.

We shall first calculate these probabilities when the parameters which
${\mathcal F}$ depends on are known i.e.\ when the filters $h_{li}$ are known
functions of time.  We shall then explain how to obtain approximate formulae for
the false alarm and the detection probabilities when parameters of the filters
are unknown.

\subsection{Detection statistics}

We shall first consider the probability density function of the normalized 
reduced functional ${\mathcal F}_1$.  Let us suppose that filters $h_{1i}$ are 
known functions of time, i.e.\ the phase parameters  
$f_0$, ${\ssbii k}$, $\alpha$, $\delta$ are known,  
and let us define the following random variables
\be
x_{1i} := (x||h_{1i}),\quad i=1,\ldots,4.
\ee
Since $x$ is a Gaussian random process the random variables $x_{1i}$ being 
linear in $x$ are also Gaussian. Let ${\rm E}_0\{x_{1i}\}$ and ${\rm 
E}_1\{x_{1i}\}$ be respectively the means  of $x_{1i}$ when the signal is absent 
and when the signal is present. One easily gets
\bea
{\rm E}_0\{x_{1i}\} = 0,\quad i=1,\ldots,4,
\eea
and
\bea
m_{11} &:=& {\rm E}_1\{x_{11}\} = \frac{1}{2}(A A_{11} + C A_{12}),\\
m_{12} &:=& {\rm E}_1\{x_{12}\} = \frac{1}{2}(C A_{11} + B A_{12}),\\
m_{13} &:=& {\rm E}_1\{x_{13}\} = \frac{1}{2}(A A_{13} + C A_{14}),\\
m_{14} &:=& {\rm E}_1\{x_{14}\} = \frac{1}{2}(C A_{13} + B A_{14}).
\eea
One finds that the covariance matrix for the random variables $x_{1i}$ is the
same whether the signal is present or not and it splits into two identical 2 by
2 covariance matrices $\mathcal{C}$ for the pairs $(x_{11},x_{12})$ and
$(x_{13},x_{14})$ of random variables where $\mathcal{C}$ is given by Eq.\
(\ref{C}).  Hence the covariance matrix is exactly equal to the matrix
$\mathcal{M}$ given by Eq.\ (\ref{M}) above.

Thus in effect $(x_{11},x_{12})$ and $(x_{13},x_{14})$ are pairs of correlated
Gaussian random variables, with pairs being independent of each other.  For
Gaussian variables their first two moments determine uniquely their probability
density function (pdf).  Consequently the joint probability density function
$p(x_{11},x_{12},x_{13},x_{14})$ is equal to a product
$p_a(x_{11},x_{12})\,p_b(x_{13},x_{14})$, where $p_a$ and $p_b$ are bivariate
Gaussian pdfs with the same covariance matrix $\mathcal{C}$:
\bea
p_a(x_{11},x_{12}) = \frac{1}{\pi \sqrt{D}} 
\exp\left( - \frac{B \tilde{x}^2_{11} + A \tilde{x}^2_{12} - 
2 C \tilde{x}_{11}\tilde{x}_{12}}{D}\right),
\eea
and a similar formula for $p_b(x_{13},x_{14})$, where $\tilde{x}_{1i}=x_{1i}$ 
when the signal is absent and $\tilde{x}_{1i}=x_{1i}-m_{1i}$ when the signal is 
present. It is interesting to note that when the signal is absent the joint pdf 
$p_{0}$ is simply given by
\be
p_{0} = \frac{1}{\pi^2 D} \exp (- {\mathcal F}_1),
\ee 
where ${\mathcal F}_1$ is our optimal statistics. We want to find the pdf of 
${\mathcal F}_1$ when the signal is absent and present. We first decorrelate 
the variables $x_{1i}$. For the case of Gaussian variables this can always be 
done by means of a linear transformation. Let us consider the following 
transformation matrix ${\mathcal L}$
\bea
{\mathcal L} =
\left(\begin{array}{cc}
{\mathcal N} & {\mathcal O} \\
{\mathcal O} & {\mathcal N}
\end{array}\right),
\eea 
where ${\mathcal O}$ is a zero $2$ by $2$ matrix and ${\mathcal N}$ is given by
\be
{\mathcal N} = \left(\begin{array}{cc}
 1/\sqrt{A + C \sqrt{A/B}} & 1/\sqrt{B + C \sqrt{B/A}} \\
-1/\sqrt{A - C \sqrt{A/B}} & 1/\sqrt{B - C \sqrt{B/A}}
\end{array}\right).
\ee
Let us introduce new random variables $z_{1i}$ $(i=1,\ldots,4)$ such that
$z_{1i}=\sum_{j=1}^4{\mathcal L}_{ij}x_{1j}$. In the new variables the pdf takes 
the form
\be
p(z_{11},z_{12},z_{13},z_{14})
= \frac{1}{(2\pi)^2}
\exp\left[ -\frac{1}{2} \left(
{\tilde z}_{11}^2 + {\tilde z}_{12}^2 + {\tilde z}_{13}^2 + {\tilde z}_{14}^2
\right) \right].
\ee
Thus $z_{1i}$ are independent Gaussian random variables with unit variances.  
When the signal is absent we have ${\tilde z}_{1i}=z_{1i}$ and when 
the signal is present ${\tilde z}_{1i}=z_{1i}- m'_{1i}$, where  
$m'_{1i}=\sum_{j=1}^4{\mathcal L}_{ij}m_{1j}$. The functional ${\mathcal F}_1$ 
in the new variables is given by
\be
{\mathcal F}_1=\frac{1}{2}\left(z_{11}^2+z_{12}^2+z_{13}^2+z_{14}^2\right).
\ee
The probability density distributions of ${\mathcal F}_1$ both when the signal 
is absent and present are well known.  When the signal is absent $2{\mathcal 
F}_1$ has a $\chi^2$ distribution with 4 degrees of freedom and when signal is 
present it has a noncentral $\chi^2$ distribution with 4 degrees of freedom and
noncentrality parameter $\lambda = \sum^4_{i=1}(m'_{1i})^2$.

In exactly the same way one obtains the pdf for the normalized reduced
functional for the second component of the signal.  The second component depends
on four amplitudes $A_{2i}$.  The decorrelation is achieved by the same matrix
${\mathcal L}$.  Let us denote the four decorrelated random variables for the
second component by $z_{2i}$ and their means by $m'_{2i}$ ($i = 1,\ldots,4$).

To obtain the pdf of the statistics ${\mathcal F}$ for the
detection of the two-component signal it is convenient to introduce
the following normalized random variables $z_i^n$:
\be
\label{nor}
z^n_i = z_{1i} \sqrt{\frac{T_o}{S_h(f_0)}},\quad
z^n_{4+i} = z_{2i} \sqrt{\frac{T_o}{S_h(2f_0)}},\quad i=1,\ldots,4,
\ee
so that each random variable $z_i^n$ has a unit variance.
Consequently $2{\mathcal F} = \sum^8_{i=1}(z_i^n)^2$ 
has a $\chi^2$ distribution with 
8 degrees of freedom when the signal is absent and noncentral $\chi^2$ with 8 
degrees of freedom when signal is present. The noncentrality parameter 
$\lambda$ is given by
\be
\lambda = \frac{T_o}{S_h(f_0)}\sum^4_{i=1}m'_{1i} +
\frac{T_o}{S_h(2f_0)}\sum^4_{i=1}m'_{2i}.
\ee
After some algebra one finds that $\lambda = d^2$ where 
\be
\label{snr}
d := \sqrt{(h|h)}.
\ee
The quantity $d$ is called the {\em optimal signal-to-noise ratio}.
It is the maximum signal-to-noise ratio that can be achieved
for a signal in additive noise with the {\em linear} filter
\cite{Da}. This fact does not depend on the statistics of the noise. 

Consequently the pdfs $p_0$ and $p_1$ when respectively the signal 
is absent and present are given by
\bea
\label{p0}
p_0({\mathcal F})
&=&\frac{{\mathcal F}^3}{6}\exp(-{\mathcal F}),
\\
\label{p1}
p_1({d,\mathcal F})
&=&\frac{(2{\mathcal F})^{3/2}}{d^3} 
I_3\left(d\sqrt{2 {\mathcal F}}\right)
\exp\left(-{\mathcal F}-\frac{1}{2}d^2\right),
\eea
where $I_3$ is the modified Bessel function of the first kind and order 3.
The false alarm probability $P_F$ is the probability that ${\mathcal F}$
exceeds a certain threshold ${\mathcal F}_o$ when there is no signal.
In our case we have
\be
P_F({\mathcal F}_o) :=
\int_{{\mathcal F}_o}^{\infty}p_0({\mathcal F})\,d{\mathcal F} =
\left(1 + {\mathcal F}_o + \frac{1}{2}{\mathcal F}_o^2 +
\frac{1}{6}{\mathcal F}_o^3\right)\exp(-{\mathcal F}_o).
\label{PF}
\ee
The probability of detection $P_D$ is the probability that ${\mathcal F}$
exceeds the threshold ${\mathcal F}_o$ when the signal-to-noise ratio is
equal to $d$:
\be
P_D(d,{\mathcal F}_o) 
:= \int^{\infty}_{{\mathcal F}_o}p_1(d,{\mathcal F})\,d{\mathcal F}.
\label{PD}
\ee
Thus we see that when the noise in the detector is Gaussian and the phase
parameters are known the probability of detection of the signal depends on a
single quantity: the optimal signal-to-noise ratio $d$.  In view of its importance
we shall investigate in detail the dependence of the optimal signal-to-noise
ratio on the parameters of the signal in the next section.

Let us introduce a vector parameter 
$\theta_i^{ph} = (f_0$, ${\ssbii k}$, $\alpha$, $\delta)$
denoting the phase parameters.
When parameters $\theta_i^{ph}$ are known the optimal statistics
${\mathcal F}$ is a random variable with probability density functions given
above.  When the phase parameters are not known we can think of ${\mathcal F}$
as a multi-dimensional random process ${\mathcal F}(\theta_i^{ph})$ with
dimension equal to the number of phase parameters.  Such a process is called
{\em random field}.
For each realization $x(t)$ of the data random process the corresponding
realization of the random field is obtained by evaluating ${\mathcal F}$ for
filters $h_{li}$ with continuously varying parameters $\theta_i^{ph}$.  
For such a process we can define the autocorrelation function 
${\bf C}$ just in the same way as we define the autocorrelation function 
for a one parameter random process:
\be
{\bf C} = E[{\mathcal F}(\theta_i^{ph}){\mathcal F}(\theta_i^{'ph})].
\ee
Let us first assume that the signal is absent, i.e.\ $x(t) = n(t)$.  For many
cases of interest the autocorrelation function will tend to zero as the
differences $\Delta _i=\theta_i^{'ph} - \theta_i^{ph}$ increase.  Thus we can
divide the parameter space into elementary cells such that 
in each cell ${\bf C}$ is appreciably different from zero.  The realizations of
the random field within a cell will be correlated (dependent) whereas
realizations of the random field within each cell and outside the cell are
almost uncorrelated (independent).  Thus the number of cells covering the
parameter space estimates the number of independent samples of the random field.
For some signals the autocorrelation function will depend only on the 
differences
$\Delta_i$ and not on the absolute values of the parameters.  Then the random
field ${\mathcal F}$ is called a {\em homogeneous random field}.
In this case one can introduce the notion of the correlation hyperellipse 
as a generalization of the correlation time of a stationary process 
and estimate the area of the elementary cell by the area of
the correlation hyperellipse.  For the general case of a random field 
the number of elementary
cells $N_c$ can be estimated from Owen's formula \cite{O96,BCCS97} with an
appropriate choice of the mismatch parameter $\mu$ and for the case of a
homogeneous random field from a formula proposed by one of us
\cite{K97a}.  For the parameter values in each cell the probability distribution 
of
${\mathcal F}(\theta_i^{ph})$ can be approximated by probability $p_0({\mathcal
F})$ given by Eq.\ (\ref{p0}).  Thus the probability distribution of ${\mathcal
F}$ is given by product of $N_c$ pdfs $p_0({\mathcal F})$.  The probability that
${\mathcal F}$ does not exceed the threshold ${\mathcal F}_o$ in a given cell is
$1 - P_F({\mathcal F}_o)$, where $P_F({\mathcal F}_o)$ is given by Eq.\
(\ref{PF}).  The probablity that ${\mathcal F}$ does not exceed the threshold
${\mathcal F}_o$ in all the $N_c$ cells is $[1 - P_F({\mathcal F}_o)]^{N_c}$.
The probability $P^T_F$ that ${\mathcal F}$ exceeds ${\mathcal F}_o$ in {\em one
or more cell} is given by
\be
P^T_F({\mathcal F}_o) = 1 - [1 - P_F({\mathcal F}_o)]^{N_c}.
\label{FP}
\ee
This is the false alarm probability when the phase parameters are unknown.  When
$P_F({\mathcal F}_o)\ll 1$ and $N_c P_F({\mathcal F}_o) < 1$ we have $P^T_F
\cong N_c P_F({\mathcal F}_o)$.  When the signal is present a precise
calculation of the pdf of ${\mathcal F}$ would be very difficult because the
presence of the signal makes the data random process $x(t)$ non-stationary.  As
a first approximation we can approximate the probability of detection of the
signal when parameters are unknown by the probability of detection when the
parameters of the signal are known [given by Eq.\ (\ref{PD})].  This
approximation assumes that when the signal is present the true values of the
phase parameters fall within the cell where ${\mathcal F}$ has a maximum.
This approximation will be the better the higher the signal-to-noise ratio $d$.
An accurate probability of detection can be obtained by numerical simulations.
Parametric plot of probability of detection vs.\ probability of false alarm with
optimal signal-to-noise ratio $d$ as a parameter is called the {\em receiver
operating characteristic} (ROC).  Detailed calculations of the number of cells
$N_c$ and false alarm probabilities as well as plots of receiver operating
characteristic for the case of the signal considered here will be given in Paper 
III.

The above reasoning is a generalization to the case of many parameters of the
idea of an {\em effective sampling rate} introduced by one of us \cite{S91} and
further developed in \cite{DSch}.  Related ideas can also be found in Ref.\
\cite{H68}.

For large signal-to-noise ratios, the rms errors of the estimators of the 
parameters of the signal are approximately given by the square roots of the 
diagonal elements of the inverse of the Fisher information matrix $\Gamma$ with 
the components given by
\be
\Gamma_{ij} =
\left(\frac{\partial{h}}{\partial{\theta_i}}\Bigg\vert
\frac{\partial{h}}{\partial{\theta_j}}\right).
\ee
We shall study these errors in detail in Paper II. For smaller signal-to-noise 
ratios (e.g.\ $\lesssim$ 10) the errors are larger (see Ref. \cite{VN}
for a discussion in the context of coalescing binaries).

\subsection{Signal-to-noise ratio}

In this subsection we use the following models of the noise spectral densities
$S_h$ in the individuals detectors.  The noise curves for the VIRGO and the
initial/advanced LIGO detectors are taken from \cite{S96}, and the noise curve
for the TAMA300 detector is taken from \cite{TAMA300,SKpc}.  Wideband and 
narrowband
versions of the GEO600 detector noise are based on \cite{KSpc}.

The optimal signal-to-noise ratio $d$ is given by the formula (\ref{snr}):
\be
\label{snr1}
d := \sqrt{(h|h)}.
\ee
The gravitational-wave signal defined by Eqs.\ (\ref{s1})--(\ref{s4}) consists 
of two narrowband components around the frequencies $f_0$ and $2 f_0$ 
and therefore to a very good accuracy the signal-to-noise ratio (\ref{snr1}) 
for that signal can be written as
\be
\label{snr3}
d \cong \sqrt{d_1^2+d_2^2},
\ee
where $d_1$ and $d_2$ are the signal-to-noise ratios for the individual
components of the signal. They are given by
\bea
\label{snr41}
d_1 &:=& \sqrt{(h_1|h_1)} \cong
\left\{ \frac{2}{S_h(f_0)}
\int_{-T_o/2}^{T_o/2}\left[h_1(t)\right]^2 dt \right\}^{1/2},\\
\label{snr42}
d_2 &:=& \sqrt{(h_2|h_2)} \cong
\left\{ \frac{2}{S_h(2f_0)}
\int_{-T_o/2}^{T_o/2}\left[h_2(t)\right]^2 dt \right\}^{1/2}.
\eea
We substitute Eqs.\ (\ref{s1})--(\ref{s4}) to Eqs.\ (\ref{snr41})--(\ref{snr42})
and drop out terms which oscillate around zero with multiples of the frequency 
$f_0$. We obtain
\bea
\label{snr51}
d_1^2&\cong&
\left[
\frac{1}{64}\sin^2 2\iota\int_{-T_o/2}^{T_o/2}F_+^2 dt
+\frac{1}{16}\sin^2 \iota\int_{-T_o/2}^{T_o/2}F_\times^2 
dt\right]\frac{h_o^2\sin^2 2\theta}{S_h(f_0)},
\\
\label{snr52}
d_2^2&\cong&
\left[
\frac{1}{4}\left(1+\cos^2\iota\right)^2\int_{-T_o/2}^{T_o/2}F_+^2 dt
+\cos^2\iota \int_{-T_o/2}^{T_o/2}F_\times^2 
dt\right]\frac{h_o^2\sin^4\theta}{S_h(2f_0)}.
\eea
After performing integrations in (\ref{snr51}) and (\ref{snr52}) we get:
\bea
\label{snr61}
d_1^2 &\cong&
\left[A_1\left(\delta,\psi,\iota\right)T_o
+B_1\left(\alpha,\delta,\psi,\iota;T_o\right)\right]
\sin^2\zeta\frac{h_o^2\sin^2 2\theta}{S_h(f_0)},\\
\label{snr62}
d_2^2 &\cong&
\left[A_2\left(\delta,\psi,\iota\right)T_o
+B_2\left(\alpha,\delta,\psi,\iota;T_o\right)\right]
\sin^2\zeta\frac{h_o^2\sin^4\theta}{S_h(2f_0)}.
\eea
The functions $B_1$ and $B_2$ are periodic in the observation time $T_o$ with
the period of two sidereal days (cf.\ Eq.\ (\ref{a2}) from Appendix B).
For simplicity we suppress the explicit dependence of the functions
$A_k$ and $B_k$ on the angles $\lambda$ and $\gamma$.
Detailed expressions for the functions $A_k$ and $B_k$ are given in Appendix B.

For the observation times $T_o$ longer than several days the signal-to-noise
ratios $d_1$, $d_2$, and $d$ are dominated by terms proportional to the square 
root of the observation time $T_o$.  This can be seen in Figures 2 and 3.

The signal-to-noise ratios $d_1^2$ and  $d_2^2$ are complicated functions of
the angles $\alpha$, $\delta$, $\psi$, $\iota$, and $\theta$.  We have studied 
the different averages of $d_1^2$ and $d_2^2$ over these angles. Averaging is 
performed according to the definition:
\be
\label{av}
\langle\cdots\rangle_{\alpha,\delta,\psi,\iota, \theta} := 
\frac{1}{2\pi}\int_0^{2\pi}d\alpha\times 
\frac{1}{2}\int_{-1}^{1}d\sin\delta \times\frac{1}{2\pi}\int_0^{2\pi}d\psi 
\times\frac{1}{2}\int_{-1}^{1}d\cos \iota 
\times\frac{1}{\pi}\int_{0}^{\pi}d\theta \left(\cdots\right).
\ee
Note that because $\delta\in[-\pi/2,\pi/2]$ integration over $\sin\delta$ 
rather than $\cos\delta$ is involved in Eq.\ (\ref{av}).

Averaging over the angle $\alpha$ discards the oscillatory parts $B_1$ and $B_2$ 
of the signal-to-noise ratios $d_1^2$ and $d_2^2$:
\bea
\label{snr71}
\langle d_1^2\rangle_{\alpha}
&\cong&A_1\left(\delta,\psi,\iota\right)\sin^2\zeta
\frac{h_o^2 T_o\sin^2 2\theta}{S_h(f_0)},
\\
\label{snr72}
\langle d_2^2\rangle_{\alpha}
&\cong&A_2\left(\delta,\psi,\iota\right)\sin^2\zeta
\frac{h_o^2 T_o\sin^4\theta}{S_h(2f_0)}.
\eea
Further averaging over the orientation angles $\psi$ and $\iota$ gives
\bea
\label{snr81}
\langle d_1^2\rangle_{\alpha,\psi,\iota}
&\cong& \frac{1}{20}e_2(\delta)\sin^2\zeta
\frac{h_o^2 T_o\sin^2 2\theta}{S_h(f_0)},
\\
\label{snr82}
\langle d_2^2\rangle_{\alpha,\psi,\iota}
&\cong& \frac{4}{5}e_2(\delta)\sin^2\zeta
\frac{h_o^2 T_o\sin^4\theta}{S_h(2f_0)}.
\eea
The function $e_2$ (its definition can be found in Appendix B) in the above
equations is a fair representation of the average sensitivity of a detector at a
given location.  It depends on the declination $\delta$ of the
gravitational-wave source as well as on the latitude $\lambda$ of the detector's
site and the angle $\gamma$ describing the orientation of its arms.  The product
$e_2(\delta)\sin^2\zeta$ is plotted against the declination $\delta$ for
different detectors in Figure 4.

Averaging over the angles $\alpha$, $\delta$, $\psi$, and $\iota$ yields 
results which do not depend on the position of the detector on the 
Earth and on the orientation of its arms:
\bea
\label{snr8a1}
\langle d_1^2\rangle_{\alpha,\delta,\psi,\iota} &\cong&
\frac{1}{100}\sin^2\zeta\frac{h_o^2T_o\sin^2 2\theta}{S_h(f_0)},
\\
\label{snr8a2}
\langle d_2^2\rangle_{\alpha,\delta,\psi,\iota} &\cong&
\frac{4}{25}\sin^2\zeta\frac{h_o^2T_o\sin^4\theta}{S_h(2f_0)}.
\eea
For the special case of the model of neutron star as a triaxial ellipsoid the 
angle $\theta = \pi/2$ and then the contribution $d^2_1$ to the signal-to-noise 
ratio vanishes. However for small angles $\theta$ the term $d^2_1$ may dominate
over the term $d^2_2$. The averaging of the above formulae over the angle 
$\theta$ gives
\bea
\label{snr99}
\langle d_1^2\rangle_{\alpha,\delta,\psi,\iota, \theta} &\cong&
\frac{1}{200}\sin^2\zeta\frac{h_o^2T_o}{S_h(f_0)},
\\
\label{snr100}
\langle d_2^2\rangle_{\alpha,\delta,\psi,\iota, \theta} &\cong&
\frac{3}{50}\sin^2\zeta\frac{h_o^2T_o}{S_h(2f_0)}.
\eea
We observe that when the noise spectral density at frequencies $f_0$ and $2f_0$ 
is the same the average (\ref{snr100}) of $d^2_2$ is more than one order of 
magnitude greater than the average (\ref{snr99}) of $d^2_1$.

We have studied the distribution of the signal-to-noise ratios $d_1$, $d_2$, and
$d$ over the angles $\alpha$, $\delta$, $\psi$, $\iota$, and $\theta$ with the
aid of the Monte Carlo simulations for the observation time $T_o = 120$ days.
For each case we have generated 10000 sets of angles according to the
probability measure defined by the right-hand side of Eq.\ (\ref{av}).  We have
assumed that the parameter $d_o$ given by Eq.\ (\ref{do}) is equal to 1.  The
results are shown in Figures 5--7 where we have plotted cumulative distribution
functions of the simulated signal-to-noise ratios $d_1$, $d_2$, and $d$ for the
initial/advanced Hanford, initial/advanced Livingston, VIRGO, GEO600, and
TAMA300 detectors.  We have performed simulations for two gravitational wave
frequencies $f_0$:  100 Hz and 500 Hz.  The shapes of the distributions of the
signal-to-noise ratios $d_1$ and $d_2$ do not depend on the frequency $f_0$
(cf.\ Eqs.\ (\ref{snr41})--(\ref{snr42})) and will be the same for
nonaxisymmetries generated by different physical mechanism e.g.\ for the case of
CFS instability.

In Table 2 we have given the means and the quartiles for the Monte Carlo
simulated cumulative distribution functions of the signal-to-noise ratios $d_1$,
$d_2$, and $d$ for the individual detectors.

From Figures 5, 6, and Table 2 we see that the simulated distributions of the 
signal-to-noise ratios $d_1$, $d_2$, and $d$ depend weakly on the position of 
the detector on the Earth and on the orientation of its arms (cf.\ plots and 
data for the initial/advanced Hanford and Livingston detectors). This is related 
to the fact that the averages (\ref{snr8a1}) and (\ref{snr8a2}) are idependent  
of the position of the detector on the Earth and of the orientation of its arms.

\begin{table}
\begin{center}
\begin{tabular}{|c|c|c|c|c|c|c|c|c|c|}\hline
detector &&
\multicolumn{4}{|c}{$f_0=100$ Hz}&
\multicolumn{4}{|c|}{$f_0=500$ Hz}\\ \cline{3-10}
&& mean & $q_{0.25}$ & $q_{0.5}$ & $q_{0.75}$ &
   mean & $q_{0.25}$ & $q_{0.5}$ & $q_{0.75}$ \\ \hline\hline
GEO600
& $d_1$ & 0.70 & 0.37 & 0.72 & 1.0
        & 14.  & 7.5  & 15.  & 21. \\ \cline{2-10}wideband noise
& $d_2$ & 3.1  & 0.79 & 2.7  & 4.7
        & 21.  & 5.4  & 18.  & 32. \\ \cline{2-10}
& $d$   & 3.2  & 1.1  & 2.9  & 4.7
        & 28.  & 17.  & 28.  & 37. \\ \hline\hline
GEO600
& $d_1$ & --   & --   & --   & --
        & 1.2  & 0.64 & 1.3  & 1.8 \\ \cline{2-10}narrowband noise
& $d_2$ & --   & --   & --   & --
        & 180. & 47.  & 160. & 280. \\ \cline{2-10}
& $d$   & --   & --   & --   & --
        & 180. & 47.  & 160. & 280. \\ \hline\hline
initial Hanford
& $d_1$ &  2.8 &  1.5 &  2.9 &   4.2
        & 48.  & 25.  & 49.  &  70. \\ \cline{2-10}
& $d_2$ & 12.  &  3.2 & 11.  &  19.
        & 72.  & 19.  & 63.  & 110. \\ \cline{2-10}
& $d$   & 13.  &  4.7 & 12.  &  19.
        & 95.  & 56.  & 96.  & 120. \\ \hline\hline
advanced Hanford
& $d_1$ &  89. &  46. &  90. &  130.
        & 480. & 250. & 490. &  700. \\ \cline{2-10}
& $d_2$ & 140. &  37. & 120. &  210.
        & 720. & 190. & 630. & 1100. \\ \cline{2-10}
& $d$   & 180. & 100. & 180. &  240.
        & 950. & 560. & 960. & 1200. \\ \hline\hline
initial Livingston
& $d_1$ & 2.9 & 1.5 & 3.0 & 4.3
        & 48. & 25. & 50. & 71. \\ \cline{2-10}
& $d_2$ & 12. & 3.2 & 11. & 19.
        & 72. & 19  & 64. & 110. \\ \cline{2-10}
& $d$   & 13. & 4.8 & 12. & 19.
        & 95. & 58. & 97. & 120. \\ \hline\hline
advanced Livingston
& $d_1$ &  89. &  47. &  92. &  130.
        & 480. & 260. & 500. &  720. \\ \cline{2-10}
& $d_2$ & 140. &  37. & 130. &  220.
        & 720. & 190. & 640. & 1100. \\ \cline{2-10}
& $d$   & 180. & 110. & 180. &  240.
        & 950. & 580. & 970. & 1200. \\ \hline\hline
VIRGO
& $d_1$ & 1.5  & 0.78 & 1.5  & 2.2
        & 46.  & 24.  & 48.  & 68. \\ \cline{2-10}
& $d_2$ & 5.8  & 1.5  & 5.2  & 8.9
        & 86.  & 22.  & 76.  & 130. \\ \cline{2-10}
& $d$   & 6.2  & 2.3  & 5.7  & 9.0
        & 110. & 57.  & 100. & 140. \\ \hline\hline
TAMA300
& $d_1$ &  0.094 & 0.049 & 0.098 & 0.14
        & 13.    & 6.9   & 14.   & 19. \\ \cline{2-10}
& $d_2$ &  1.1   & 0.28  & 0.97  & 1.7
        & 29.    & 7.3   & 25.   & 44. \\ \cline{2-10}
& $d$   &  1.1   & 0.30  & 0.98  & 1.7
        & 34.    & 17.   & 33.   & 46. \\ \hline\hline
initial LIGO/VIRGO
& $d_1$ & 4.3  & 2.3  & 4.5  & 6.4
        & 82.  & 43.  & 86.  & 120. \\ \cline{2-10}network
& $d_2$ & 19.  & 4.8  & 17.  & 28.
        & 130. & 35.  & 120. & 200. \\ \cline{2-10}
& $d$   & 20.  & 7.1  & 18.  & 29.
        & 170. & 100. & 170. & 230. \\ \hline
\end{tabular}
\end{center}
\caption{The means and the quartiles for the Monte Carlo simulated distribution 
functions of the signal-to-noise ratios $d_1$, $d_2$, and $d$ for the individual 
detectors and for the three detector network of the VIRGO and two initial LIGO 
detectors. For the GEO600 detector we use two noise curves: wideband and 
narrowband tuned to 1 kHz with the bandwidth of 30 Hz.  We assume that star's 
ellipticity is $10^{-5}$, its moment of inertia w.r.t.\ the rotation axis 
is $10^{45}$ g cm$^2$ and its distance from the Earth is 1 kpc. Quantile $q_x$ 
gives a value $z_x$ of random variable $z$ such that probability that $z<z_x$ is 
less than or equal to $x$. The quantile values at $x$ = 0.25, 0.5, and 0.75 are 
called the quartiles.}
\end{table}

\subsection{Data analysis method}

It is important to calculate the optimum statistics as efficiently as possible.
One way to achieve this is to take advantage of the speed of the {\em fast
Fourier transform} (FFT).  Let us consider first the normalized reduced
functional ${\mathcal F}_1$.  One observes that the phase $\Phi$ of the signal
can be written as (cf.\ Eq.\ (\ref{pha6b}))
\be
\Phi(t) = 2\pi f_0 [t + \Phi_m(t;\alpha,\delta)]
+ \Phi_s(t;{\ssbii k},\alpha,\delta),     
\ee
where functions $\Phi_m$ and $\Phi_s$ do not depend on the frequency parameter 
$f_0$. Let us define the following two integrals:
\bea
F_{1a} &=& \int^{T_o/2}_{-T_o/2} x(t)a(t)
\exp[-i\Phi_s(t)]
\exp\left\{-i 2\pi f_0 [t + \Phi_m(t)]\right\}\,dt,
\\
F_{1b} &=& \int^{T_o/2}_{-T_o/2} x(t)b(t)
\exp[-i\Phi_s(t)]
\exp\left\{-i 2\pi f_0 [t + \Phi_m(t)]\right\}\,dt.
\eea
One can write the statistics ${\mathcal F}_1$ in terms of the above two 
integrals as:
\be
{\mathcal F}_1 = \frac{4}{T_o^2}\frac{B |F_{1a}|^2 + A |F_{1b}|^2 - 
2 C \Re(F_{1a} F_{1b}^*)}{D}.
\ee
We can introduce a new time coordinate $t_b$:
\be
\label{Rs}
t_b(t) = t + \Phi_m(t).
\ee
From the explicit expression for the phase $\Phi$ given by Eq.\ (\ref{pha6b})
the time shift $\Phi_m$ and its time derivative $\dot{\Phi}_m$ can be estimated 
by
\be
\begin{array}{rcccl}
\vert\Phi_m(t)\vert &\lesssim&
{\dst \frac{R_{ES}}{c}} &\simeq& 5\times 10^2\,\,{\rm s},
\\[2ex]
\vert\dot{\Phi}_m(t)\vert &\lesssim&
{\dst \frac{\Omega_o R_{ES}}{c}} &\simeq& 1\times 10^{-4}.
\end{array} 
\ee
Assuming the maximum observation time $T_o=120$ days to a very good 
approximation we have
\be
T_b := t_b(T_o) \cong T_o,\quad \frac{dt}{dt_b} \cong 1.
\ee
Thus in the new time coordinate the integrals $F_{1a}$ and $F_{1b}$ can be very 
well approximated by
\bea
\label{i1a}
F_{1a} &\cong& \int^{T_o/2}_{-T_o/2} x[t(t_b)] a[t(t_b)] 
\exp\{-i\Phi_s[t(t_b)]\}
\exp(-i2\pi f_0 t_b)\,dt_b,
\\
\label{i1b}
F_{1b} &\cong& \int^{T_o/2}_{-T_o/2} x[t(t_b)] b[t(t_b)] 
\exp\{-i\Phi_s[t(t_b)]\}
\exp(-i2\pi f_0 t_b)\,dt_b.
\eea
Hence we see that with the new time coordinate $t_b$ the two integrals
(\ref{i1a}) and (\ref{i1b}) are Fourier transforms of the functions $x[t(t_b)]
a[t(t_b)]\exp\{-i\Phi_s[t(t_b)]\}$ and $x[t(t_b)]
b[t(t_b)]\exp\{-i\Phi_s[t(t_b)]\}$, respectively.  
To calculate these integrals for a given set of phase parameters we need to
perform the following numerical operations.
For the chosen values of the parameters $\alpha$ and $\delta$ we resample
the original time series according to the formula (\ref{Rs}) and then we
multiply the resampled time series $x(t_b)$ by functions
$a(t_b)\exp\{-i\Phi_s[t(t_b)]\}$ and $b(t_b)\exp\{-i\Phi_s[t(t_b)]\}$.  Then we
calculate the two Fourier transforms (using FFT algorithm).
The resampling technique has been proposed by one of us
\cite{S91} and considered as an effective data analysis tool for searches of
gravitational waves from periodic sources \cite{BCCS97}.

Alternatively one could define new spindown parameters
\be
f_k := \frac{{\ssbii k}}{f_0},\quad k=1,\ldots,s,
\ee
and introduce a different time coordinate
\be
\label{tb'}
t'_{b}(t) = t + \Phi_m(t) + \Phi'_{s}(t),
\ee
where
\be
\Phi'_{s}(t) = \sum_{k=1}^{s}f_k\frac{t^{k+1}}{(k+1)!}
\ee
and perform the resampling process according to the formula (\ref{tb'}).

The functions $a$, $b$ and consequently $A$, $B$, $C$, and $D$ are known and
they depend on the declination, the right ascension and the time of observation.
Their values can be calculated and stored for a fine grid of positions of the
neutron star on the sky and appropriate observation times before the data 
analysis is
carried out.

The normalized reduced functional for the second component of the signal can be 
calculated in a similar way. Here the corresponding Fourier transforms are given 
by
\bea
F_{2a} &\cong& \int^{T_o/2}_{-T_o/2} x[t(t_b)] a[t(t_b)] 
\exp\{-i2\Phi_s[t(t_b)]\}
\exp(-i4\pi f_0 t_b)\,dt_b,
\\
F_{2b} &\cong& \int^{T_o/2}_{-T_o/2} x[t(t_b)] b[t(t_b)] 
\exp\{-i2\Phi_s[t(t_b)]\}
\exp(-i4\pi f_0 t_b)\,dt_b.
\eea
The statistics ${\mathcal F}$ for the whole signal is then calculated from the 
formula
\bea
{\mathcal F} &=& 
\frac{4}{S_h(f_0)T_o} \frac{B |F_{1a}|^2 + A |F_{1b}|^2 - 
2 C \Re(F_{1a} F_{1b}^{*})}{D}
\nonumber\\&&
+ \frac{4}{S_h(2f_0)T_o}\frac{B |F_{2a}|^2 + A |F_{2b}|^2 - 
2 C \Re(F_{2a} F_{2a}^{*})}{D}.
\eea
The statistics ${\mathcal F}$ needs to be calculated on a multidimensional grid
of parameter values (excluding the frequency parameter $f_0$) covering
sufficiently densely the parameter space, and compared against a threshold.

\section{Networks of detectors}

The analysis of the previous section can be generalized to the case of a network
of $N$ interferometers in a straightforward manner.  Assuming that the noise in
each detector is uncorrelated with the others, the likelihood
function for the network is the sum of the likelihood functions for the
individual detectors.  Therefore we define a statistics ${\mathcal F}_n$ for the
whole network as the sum of the individual statistics of each detector given by
Eq.\ (\ref{STAT}).  We maximize ${\mathcal F}_n$ with respect to the phase
parameters to obtain their estimators.  We calculate the estimators of the
amplitudes from the analytic fomulae.  Then we use a least-squares fit to
estimate the parameters $(h_o,\theta,\psi,\iota,\Phi_0)$ from the 8$N$
amplitude estimators.  When the phase parameters of the signal are known each of
the individual statistics ${\mathcal F}_i$ multiplied by a factor of 2 has
$\chi^2$ probability density distribution with 8 degrees of freedom when the
signal is absent and noncental $\chi^2$ distribution with noncentrality
parameter $d^2_i$ when the signal is present.  The Gaussian variables entering
each statistics (normalized random variables $z^n_i$ given by Eqs.\ (\ref{nor}))
have the same unit variance.  Thus $2{\mathcal F}_n$ has the $\chi^2$
distribution with $8 N$ degrees of freedom when signal is absent and noncentral
$\chi^2$ distribution with noncentrality parameter $d_n^2 = \sum^N_{i=1} d^2_i$
when the signal is present.  The quantity $d_n$ can be defined as the {\em total
signal-to-noise ratio} of the network.  Probability of detection is then
caclulated by Eq.\ (\ref{PD}).  When the phase parameters of the signal are
unknown similarly like in the one-detector case one can consider a random field
which is a sum of the random fields for the individual detectors and investigate
the correlation function for this random field to obtain independent number of
cells $N_c$ of the field.  One can then calculate the false alarm probability
for the network by means of Eq.\ (\ref{FP}).

We have studied the distribution of the network signal-to-noise ratios $d_{n1}$,
$d_{n2}$, and the total network signal-to-noise ratio
$d_n\cong\sqrt{d_{n1}^2+d_{n2}^2}$ over the angles $\alpha$, $\delta$, $\psi$,
$\iota$, and $\theta$ with the aid of the Monte Carlo simulations for the
observation time $T_o = 120$ days.  We have restricted ourselves to the three
detector network of the VIRGO and two initial LIGO detectors.  For each case we
have generated 10000 sets of angles according to the probability measure defined
by the right-hand side of Eq.\ (\ref{av}).  We have assumed that the parameter
$d_o$ given by Eq.\ (\ref{do}) is equal to 1.  The results are shown in Figure
8.  We have performed simulations for two gravitational wave frequencies $f_0$:
100 Hz and 500 Hz.

In Table 2 we have put the means and the quartiles for the Monte Carlo simulated
cumulative distribution functions of the signal-to-noise ratios $d_{n1}$,
$d_{n2}$, and $d_n$ for the three detector network of the VIRGO and two initial
LIGO detectors.  Adding the GEO600 and TAMA300 detectors will not
significantly change these signal-to-noise ratio values, but the smaller
detectors can play an important role in making coincident detections by
improving the confidence that the candidate events registered by larger
detectors are not due to un-modelled noise.

\section*{Acknowledgments}

Two of us (P.J.\ and A.K.)  would like to thank the Albert Einstein Institute,
Max Planck Institute for Gravitational Physics for hospitality during the visits 
when this work was done.  We also thank C.\ Cutler and M.\ Tinto for helpful 
discussions.


\appendix

\section{The phase of the gravitational-wave signal}

We assume that in the rest frame of the neutron star 
the time dependence of the phase of the gravitational-wave signal
can be written  as a power series of the 
form:
\be
\label{apa1}
\Psi_{\rm ns}(\tau) = \Phi_0
+ 2\pi\sum_{k=0}^{s}{\nssd k}\frac{\tau^{k+1}}{(k+1)!},
\ee
where $\tau$ is the proper time in the neutron star rest frame.
The assumption (\ref{apa1}) means that the instantaneous frequency of the signal 
in the rest frame of the neutron star is given as
\be
\label{apa2}
f_{\rm ns}(\tau)
:= \frac{1}{2\pi} \frac{d\Psi_{\rm ns}(\tau)}{d\tau}
=\sum_{k=0}^{s}{\nssd k}\frac{\tau^k}{k!},
\ee
so ${\nssd k}$ is the $k$th time derivative of the frequency 
evaluated at $\tau=0$.

We assume that the neutron star is moving with respect to the SSB uniformly 
along a straight line according to the equation
\be
\label{apa3}
{\bf r}_{\rm ns}(t)=r_0 {\bf n}_0+v_{\rm ns} {\bf n}_v 
\left(t+\frac{r_0}{c}\right),
\ee
where $r_0:=\vert{\bf r}_{\rm ns}(t=-r_0/c)\vert$,
${\bf n}_0:={\bf r}_{\rm ns}(t=-r_0/c)/r_0$. If we denote by
${\bf v}_{\rm ns}$ the constant velocity vector of the neutron star then
$v_{\rm ns}:=\vert{\bf v}_{\rm ns}\vert$ and
${\bf n}_v:={\bf v}_{\rm ns}/v_{\rm ns}$. 
The time $t$ in Eq.\ (\ref{apa3}) is the time 
coordinate in the SSB rest frame. We do not allow the neutron star to have an 
intrinsic acceleration. This means we exclude binary neutron stars, except of 
the binary periods so long that the acceleration effects may be accurately 
approximated by a Taylor series during the observation time.

The phase observed at the SSB at some time 
$t$ was emitted by the star at the coordinate time $t'$ such that
\be
\label{apa4}
t = t' + \frac{\vert{\bf r}_{\rm ns}(t')\vert}{c}.
\ee
One can show that the relation between the time $t'$ 
and the star's proper time $\tau$ 
is as follows
\be
\label{apa5}
\tau = \sqrt{1-\bns^2} \left(t'+\frac{r_0}{c}\right),
\ee
where $\bns:=v_{\rm ns}/c$. In Eq.\ (\ref{apa5}) the time dilation effect is 
taken 
into account. We have also assumed that $\tau=0$ when the star's position vector 
w.r.t.\ the SSB is $r_0 {\bf n}_0$. We can now write
\be
\label{apa6}
\Psi_{\rm SSB}(t) = \Psi_{\rm ns}(\tau(t)),
\ee
where $\Psi_{\rm SSB}(t)$ is the phase observed at the SSB at time $t$, and the 
time 
$\tau$ can be expressed in terms of $t$ by means of Eqs.\ (\ref{apa4}) and 
(\ref{apa5}).

Collecting Eqs.\ (\ref{apa1}) and (\ref{apa4})--(\ref{apa6}) together we can 
write
\be
\label{apa7}
\Psi_{\rm SSB}(t) = \Phi_0
+ 2\pi\sum_{k=0}^{s}\frac{{\nssd k}}{(k+1)!} 
\left(1-\bns^2\right)^{(k+1)/2} \left(t'(t)+\frac{r_0}{c}\right)^{k+1},
\ee
where $t'$ is the solution of Eq.\ (\ref{apa4}) for a given time $t$. It reads
\bea
\label{apa8}
t' &=& \frac{1}{1-\bns^2}\Bigg\{t
+ \frac{r_0}{c}\bns \left[\bns+\left({\bf n}_0\cdot{\bf n}_v\right)\right]
\nonumber\\&&
- \sqrt{\bns^2 t^2
+ 2\frac{r_0}{c}\bns \left[\bns+\left({\bf n}_0\cdot{\bf n}_v\right)\right] t
+ \frac{r_0^2}{c^2} \left[1 + \bns\left({\bf n}_0\cdot{\bf n}_v\right)\right]}
\Bigg\}.
\eea
We expand the function $\Psi_{\rm SSB}$ given by Eqs.\ (\ref{apa7}) and 
(\ref{apa8}) w.r.t.\ time 
$t$ around $t=0$. The first few terms of the expansion read
\bea
\label{apa9}
\frac{\Psi_{\rm SSB}(t) - \Phi_0}{2\pi} &=& {\ssbi 0}\,t
+ \bigg\{ {\ssbi 1} + 
\frac{\left(\nonv^2-1\right)\bns^2}{(1+\nonv\bns)^2(r_0/c)} {\ssbi 0}
\bigg\} \frac{t^2}{2}
\nonumber\\&&\kern-25ex
+ \bigg\{ {\ssbi 2}
+ \frac{3\left(\nonv^2-1\right)\bns^2}{\left(1+\nonv\bns\right)^2(r_0/c)}
\bigg[{\ssbi 1}
- \frac{(\bns+\nonv)\bns}{\left(1+\nonv\bns\right)^2(r_0/c)} {\ssbi 0} \bigg]
\bigg\} \frac{t^3}{6} + {\mathcal O}(t^4),
\eea
where
\be
\label{apa10}
{\ssbi k} := \frac{\left(1-\bns^2\right)^{(k+1)/2}}
{\left(1+\left({\bf n}_0\cdot{\bf n}_v\right)\bns\right)^{k+1}}
{\nssd k},\quad k=0,\ldots,s.
\ee
As a result of the motion of the neutron star w.r.t.\ the SSB the Taylor 
expansion (\ref{apa9}) of the phase $\Psi_{\rm SSB}$ contains infinitely many 
terms, even if we restrict, as in Eq.\  (\ref{apa1}), the intrinsic spindown of 
the star to finite number of terms. When the neutron star moves radially w.r.t.\ 
the SSB then  $\left({\bf n}_0\cdot{\bf n}_v\right)^2=1$ and the function 
$\Psi_{\rm SSB}$ can {\em exactly} be written as the finite sum:
\be
\label{apa11}
\Psi_{\rm SSB}(t) = \Phi_0
+ 2\pi\sum_{k=0}^{s}{\ssbi k}\frac{t^{k+1}}{(k+1)!}.
\ee

We shall assume the following polynomial model of the phase of the gravitational 
radiation observed at the SSB: 
\be
\label{apa12}
\Psi_{\rm SSB}(t) = \Phi_0
+ 2\pi\sum_{k=0}^{s}{\ssbii k}\frac{t^{k+1}}{(k+1)!},
\ee
where the new spindown parameters ${\ssbii k}$ do not in general coincide with 
the Doppler scaled intrinsic spindown parameters ${\ssbi k}$ defined by Eq.\ 
(\ref{apa10}). 

We write the position vector ${\bf r}_{\rm d}$ of the detector with respect to 
the SSB as
\be
\label{apa13}
{\bf r}_{\rm d}(t)=r_{\rm d}(t) {\bf n}_{\rm d}(t),
\ee
where $r_{\rm d}(t):=\vert{\bf r}_{\rm d}(t)\vert$ and
${\bf n}_{\rm d}(t):={\bf r}_{\rm d}(t)/r_{\rm d}(t)$.
The phase of the gravitational-wave signal at the time $t$ at the detector's 
location corresponds to the phase near the neutron star at an earlier instant of 
time $t''$, where $t''$ is the solution of the equation
\be
\label{apa14}
t = t'' + \frac{\vert{\bf r}_{\rm ns}(t'')-{\bf r}_{\rm d}(t)\vert}{c}.
\ee
The same value of the phase is observed at the SSB at time
$$
t'' + \frac{\vert{\bf r}_{\rm ns}(t'')\vert}{c},
$$
thus using Eq.\ (\ref{apa12}) we can write
\be
\label{apa15}
\Psi_{\rm d}(t) = \Phi_0
+ 2\pi\sum_{k=0}^{s}\frac{\ssbii k}{(k+1)!}
\left( t''(t) + \frac{\vert{\bf r}_{\rm ns}(t''(t))\vert}{c}\right)^{k+1},
\ee
where  $t''(t)$ is the solution of Eq.\ (\ref{apa14}) for given time $t$. Using 
Eqs.\ (\ref{apa3}) and (\ref{apa13}) we express the solution $t''$ to Eq.\ 
(\ref{apa14}) in terms of the time $t$ and the two small parameters $\beta_{\rm 
ns}$ and $x:=r_{\rm d}/r_0$:
\bea
\label{apa16}
t''(x,\beta_{\rm ns}) 
&=& \frac{1}{1-\beta_{\rm ns}^2}
\Bigg\{ t
+ \frac{r_0}{c}\beta_{\rm ns} \left[\nonv - \ndnv x + \beta_{\rm ns}\right]
\nonumber\\&&\kern-12ex
-\left[ \left(
t + \frac{r_0}{c}\beta_{\rm ns} \left[\nonv - \ndnv x + \beta_{\rm ns}\right]
\right)^2 + \left(1-\beta_{\rm ns}^2\right)
\right.\nonumber\\&&\left.\kern-12ex
\times \left( \frac{r_0^2}{c^2}
\left[ 1  + \bns^2 - 2 \nond x + x^2
+ 2\left(\nonv - \ndnv x\right) \bns \right] - t^2 
\right)
\right]^{1/2}\Bigg\}.
\eea
Using Eq.\ (\ref{apa13}) we also find that
\be
\label{apa17}
\frac{\vert{\bf r}_{\rm ns}(t'')\vert}{c}
= \sqrt{
\frac{r_0^2}{c^2}
+ 2\frac{r_0}{c}\left(t''+\frac{r_0}{c}\right)\nonv\bns
+ \left(t''+\frac{r_0}{c}\right)^2 \bns^2 }.
\ee
We now study how to simplify the phase $\Psi_{\rm d}$ given by Eqs.\ 
(\ref{apa15})--(\ref{apa17}).

An optimal method to detect our signal in noise developed in Section 3 involves
correlating the data with templates of the signal.  In general if the phase of
the template differs from that of the signal by as little as 1/4 of a cycle the
correlation will be significantly reduced.  Thus we adopt the criterion that
{\em we exclude an effect from the model of the signal in the case when it
contributes less than 1/4 of a cycle to the phase of the signal during the
observation time}.  That this criterion is only a sufficient condition but not
necessary follows from the correlations among parameters of the
phase.  The shifts in the values of the parameters in the template phase away
from the true values of the parameters in the signal phase can compensate for
the effects in the signal not taken into account in the templates.  This effect
was observed for the case of coalescing binaries \cite{Cal,BD,S,Kc}.  Finally we
stress that such shifts in the template parameter values mean that the
estimators of the parameters of the signal when using an inaccurate template
will be biased.  It may happen that these biases are much larger than the
rms errors of the estimators.  Thus templates accurate to 1/4 of a cycle
over the observation time may not be needed to detect the signal, but they will
be needed to obtain accurate estimates of the errors in parameter measurements.

In calculating the number of cycles we assume a long
observation time of $120$ days, the maximum gravitational wave frequency
of 1 kHz, and the extreme case of a neutron star at a distance 
$r_0 = 40$ pc with $v_{\rm ns} = 10^3$ km/s. 
For this extreme case the parameters $x$ and $\beta_{\rm ns}$ assume the values
(as to a good approximation $r_{\rm d}\cong1$ AU):
\be
\label{ap11}
x = 1.21\times10^{-7},\quad \beta_{\rm ns} = 3.34\times10^{-3}.
\ee
The numerical values of the spindown parameters 
${\ssbii k}$ we estimate by means of the relation:
\be
\bigg\vert{\ssbii k}\bigg\vert
\simeq k!\frac{f_0}{\tau^k},
\ee
where $f_0$ is the radiation frequency and $\tau$ is the spindown age of the 
neutron star. As the extreme case we will consider $\tau=$ 40 years.

It is convenient to carry out the Taylor expansion of the phase (\ref{apa15}) 
with respect to the parameters $x$ and $\beta_{\rm ns}$.
We note that for any $n$
$$
\frac{\pa^n\Psi_{\rm d}}{\pa\beta_{\rm ns}^n}\left(x=0,\bns=0\right)=0.
$$
Analysis of the first few terms of the Taylor expansion shows that for 
the observation times $T_0\le$ 120 days, neutron star distances $r_0\ge$ 40 pc, 
velocities 
$v_{\rm ns}\le$ $ 10^3$ km/s, frequencies $f_0\le$ 1 kHz, and spindown ages 
$\tau\ge$ 
40 years, the only terms which can contribute more than 1/4 of a cycle to the 
phase of the signal, read
\be
\label{ap200}
\Psi_{\rm d}\cong\Phi_0
+ 2\pi \sum_{k=0}^4 {\ssbii k}\frac{t^{k+1}}{(k+1)!}
+ \frac{2\pi}{c}\left(
{\bf n}_0 + \frac{{\bf v}_{{\rm ns}\perp}}{r_0} t
\right)\cdot{\bf r}_{\rm d}
\sum_{k=0}^3 {\ssbii k}\frac{t^k}{k!},
\ee
where
${\bf v}_{{\rm ns}\perp}
:={\bf v}_{\rm ns}-\left({\bf n}_0\cdot{\bf v}_{\rm ns}\right){\bf n}_{0}$ 
is the component of the velocity ${\bf v}_{\rm ns}$ perpendicular to the vector 
${\bf n}_{0}$. The ratio ${\bf v}_{{\rm ns}\perp}/r_0$ determines the proper 
motion of the star. The term in the above expansion proportional to
${\bf v}_{{\rm ns}\perp}/r_0$ contributes at most $\sim$4 cycles.  We shall not 
consider it in this paper.  We shall look at the possibility of its 
determination in the next paper of this series. Consequently  we restrict 
ourselves to a phase model at the detector of the form
\be
\label{ap21a}
\Psi_{\rm d} \cong \Phi_0
+ 2\pi\sum_{k=0}^{4}{\ssbii k}\frac{t^{k+1}}{(k+1)!}
+ \frac{2\pi}{c} {\bf n}_0\cdot{\bf r}_{\rm d} 
\sum_{k=0}^{3}{\ssbii k}\frac{t^k}{k!}.
\ee
The model (\ref{ap21a}) contains the position ${\bf r}_{\rm d}$ of the Earth 
relative to the SSB, which we now consider.  In addition we must consider 
extra, purely relativistic effects left out of (\ref{ap21a}).

Motion of the Earth w.r.t.\ the SSB is very well determined and there are
several computer ephemeris routines available \cite{J95}.  In this paper we
assume for simplicity that the Earth moves on a circular orbit around the Sun.
The eccentricity of the Earth's orbit ($e_\oplus = 0.017$) introduces a change
of about $8.3\times10^3$ cycles in the phase w.r.t.\ the phase for circular
orbit for 1 kHz signal, so it must be included in realistic filters.  But it
introduces no new parameters so we ignore it here.  We also ignore the motion of
the Earth around the Earth-Moon barycenter.

There are two types of relativistic corrections.  One originates in the
difference between the coordinate time $t$ which we used in the derivation of
the phase model and the proper time $\tau$ in the detector's reference frame.
The difference is due to the combined effect of the gravitational redshift and
the time dilation.  The other correction is the Shapiro delay caused by
propagation of the gravitational wave through the curved spacetime of the solar
system.  We estimate the contribution to the number of cylces in the phase
produced by these corrections.

The difference between the coordinate time $t$ in the first order post-Newtonian 
coordinate system which is assumed to be the rest frame of the SSB and the 
proper time $\tau$ kept by a terrestial clock is discussed in detail in Ref.\ 
\cite{BH86}. The difference $\Delta_E := t - \tau$ is given by the integral
\be
\label{rel1}
\Delta_E = \frac{1}{c^2}
\int\limits_0^t \left\{
U\left[{\bf r}(t')\right]
+ \frac{v(t')^2}{2} \right\}dt',
\ee
where ${\bf r}$ is the position vector of the clock w.r.t.\ the SSB, ${\bf v}
:= \dot{\bf r}$ is the clock's coordinate velocity, and $U\left[{\bf
r}(t)\right]$ is the instantaneous gravitational potential at the clock's
location.  The time difference described by the integral (\ref{rel1}) can be
split into the secular and periodic part.  The secular difference is due to the
practically constant rotational velocity and the Earth's gravitational potential
at the detector's location as well as the average orbital velocity of the Earth
and the average gravitational potential along the Earth's orbit.  This secular
difference corresponds to the rescaling of the time coordinate and can be
incorporated into the definition of the spindown parameters.  The main
contribution to the periodic part of the integral (\ref{rel1}) was calculated by
Clemence and Szebehely \cite{CS67} and then corrected by Blandford and Teukolsky
\cite{BT76}.  It can be written as
\be
\label{rel2}
\left(\Delta_E\right)_{\rm periodic}
\cong
\frac{2GM_\odot e_\oplus}{c^2 a_\oplus \left(1-e^2_\oplus\right) \Omega_o}
\left [\left(1-\frac{1}{8}e^2_\oplus\right)\sin M_\oplus
+\frac{1}{2}e_\oplus\sin 2M_\oplus
+\frac{3}{8}e^2_\oplus\sin 3M_\oplus \right],
\ee
where $M_\odot$ is the mass of the Sun, $a_\oplus$ = 1 AU, $\Omega_o$ is the
mean orbital angular velocity of the Earth, $e_\oplus$ and $M_\oplus$ are the
eccentricity and mean anomaly of the Earth's orbit.  The quantity
$\left(\Delta_E\right)_{\rm periodic}$ varies in time with the period of one
year and has the amplitude $\simeq 1.7\times10^{-3}$ s, so for a 1 kHz 
gravitational wave the contribution of this correction to the total number of 
cycles is not greater than $\sim$2 cycles.  Even when it must be included in a 
filter, it introduces no new parameters.

The magnitude of the Shapiro delay can be estimated from the relation \cite{S64} 
(neglecting the eccentricity of the Earth's orbit)
\be
\label{rel3}
\Delta_S = \frac{2GM_\odot}{c^3} \log\frac{1}{1+\cos\theta},
\ee
where $\theta$ is the star-Sun-detector 
angle at the time of observation. To estimate the maximum value of 
the Shapiro delay we consider a neutron star in such position that at 
some instant of time the line of sight from the detector to the neutron star is 
tangent to the surface of the Sun. Then $\theta=\theta_1\simeq\pi-\zeta$, where 
$\zeta\simeq R_\odot/\mbox{1 AU}\simeq 4.65\times10^{-3}$ rad ($R_\odot$ is 
the radius of the Sun). Six months later $\theta=\theta_2\simeq\zeta$, so the 
amplitude of the correction is
$$
\Delta_S\left(\theta=\theta_1\right)-\Delta_S\left(\theta=\theta_2\right)
\simeq\ \frac{2GM_\odot}{c^3} \log\frac{1+\cos\theta_2}{1+\cos\theta_1}
\simeq 1.2\times10^{-4}\,\,{\rm s}.
$$
For a 1 kHz gravitational wave this gives $\sim$0.1 cycles. So the Shapiro delay 
will be unobservable.

We see that the relativistic corrections that need to be applied to our formula
are small.  By our 1/4 of a cycle criterion they can be neglected if we search
for signals with frequencies less than $\sim$100 Hz.  We shall not consider
these corrections in this and the following papers of the series since they are
unlikely influence our results.  However they may need to be included in
filters.

\section{Signal-to-noise ratio}

In this appendix we give the detailed expressions for the functions $A_1$, 
$A_2$, $B_1$, and $B_2$ from Eqs.\ (\ref{snr61}) and (\ref{snr62}). They read:
\bea
\label{a1}
A_k\left(\delta,\psi,\iota\right)
&=& F_k(\iota)e_1(\delta)\cos4\psi
+G_k(\iota)e_2(\delta),
\\
\label{a2}
B_k\left(\alpha,\delta,\psi,\iota;T_o\right)
&=& \frac{1}{\Omega_r}\sum_{n=1}^4
\sin\left(n\frac{\Omega_r}{2}T_o\right)
\nonumber\\&&\times
\left\{
C_{kn}\left(\delta,\psi,\iota\right)\cos\left[n(\alpha-\phi_r)\right]
+D_{kn}\left(\delta,\psi,\iota\right)\sin\left[n(\alpha-\phi_r)\right]\right\},
\eea
where $\Omega_r$ is the rotational frequency of the Earth, $T_o$ is the 
observation time, and where
\bea
\label{a3}
C_{kn}\left(\delta,\psi,\iota\right)
&=& F_k(\iota)\left[f_{1n}(\delta)\cos4\psi+g_{1n}(\delta)\sin4\psi\right]
+G_k(\iota)h_{1n}(\delta),
\\
\label{a4}
D_{kn}\left(\delta,\psi,\iota\right)
&=& F_k(\iota)\left[f_{2n}(\delta)\cos4\psi+g_{2n}(\delta)\sin4\psi\right]
+G_k(\iota)h_{2n}(\delta),
\\
\label{a5}
F_1(\iota)&=&-\frac{1}{16}\sin^4\iota,\quad
F_2(\iota)=\frac{1}{4}\sin^4\iota,
\\
\label{a6}
G_1(\iota)&=&\frac{1}{16}\sin^2\iota\left(1+\cos^2\iota\right),\quad
G_2(\iota)=\frac{1}{4}\left(1+6\cos^2\iota+\cos^4\iota\right).
\eea
The functions $e_1$, $e_2$, $f_{kn}$, and $g_{kn}$ ($k=1,2$, $n=1,\ldots,4$) 
entering Eqs.\ (\ref{a1}), (\ref{a3}), and (\ref{a4}) are equal to
$$
\begin{array}{rclrcl}
e_1(\delta) &=& 4j_1 \cos^4\delta,&
e_2(\delta) &=& 4j_2 - j_3\cos2\delta + j_1\cos^22\delta,
\\
f_{11}(\delta) &=& -4j_4 \cos^3\delta\sin\delta,&
f_{12}(\delta) &=&   j_5 \cos^2\delta\left(3 - \cos2\delta\right),
\\
f_{13}(\delta) &=& -j_6 \left(7-\cos2\delta\right)\sin2\delta,&
f_{14}(\delta) &=& -j_7 \left(35-28\cos2\delta+\cos4\delta\right),
\\
f_{21}(\delta) &=& -28j_8 \cos^3\delta\sin\delta,&
f_{22}(\delta) &=&  -7j_9 \left(3 - \cos2\delta\right)\cos^2\delta,
\\
f_{23}(\delta) &=& -j_{10} \left(7 - \cos2\delta\right)\sin2\delta,&
f_{24}(\delta) &=& -j_{11} \left(35 - 28\cos2\delta+\cos4\delta\right),
\\
g_{11}(\delta) &=& 28j_8 \cos^3\delta,&
g_{12}(\delta) &=& 28j_9 \cos^2\delta\sin\delta,
\\
g_{13}(\delta) &=&  2j_{10} \left(5 - 3\cos2\delta\right)\cos\delta,&
g_{14}(\delta) &=& 16j_{11} \left(3 - \cos2\delta\right)\sin\delta,
\\
g_{21}(\delta) &=& -4j_4 \cos^3\delta,&
g_{22}(\delta) &=&  4j_5 \cos^2\delta\sin\delta,
\\
g_{23}(\delta) &=&  -2j_6 \left(5 - 3\cos2\delta\right)\cos\delta,&
g_{24}(\delta) &=& -16j_7 \left(3 - \cos2\delta\right)\sin\delta,
\\
h_{11}(\delta) &=& \left(j_{12} - j_4\cos2\delta\right)\sin2\delta,&
h_{12}(\delta) &=& \left(j_{13} - j_5\cos2\delta\right)\cos^2\delta,
\\
h_{13}(\delta)&=&  4j_6 \cos^3\delta\sin\delta,&
h_{14}(\delta)&=& -8j_7 \cos^4\delta,
\\
h_{21}(\delta) &=&  j_8 \left(1 - 7\cos2\delta\right)\sin2\delta,&
h_{22}(\delta) &=& -j_9 \left(5 - 7\cos2\delta\right)\cos^2\delta,
\\
h_{23}(\delta) &=&  4j_{10} \cos^3\delta\sin\delta,&
h_{24}(\delta) &=& -8j_{11} \cos^4\delta,
\end{array}
$$
where the coefficients $j_1,\ldots,j_{13}$ depend on the angles $\lambda$ and 
$\gamma$:
\ben
j_1\left(\lambda,\gamma\right)
&=& \frac{1}{256} \left(
4 - 20\cos^2\lambda + 35\sin^22\gamma\cos^4\lambda \right),
\\
j_2\left(\lambda,\gamma\right)
&=& \frac{1}{1024} \left(
68 - 20\cos^2\lambda - 13\sin^22\gamma\cos^4\lambda \right),
\\
j_3\left(\lambda,\gamma\right)
&=& \frac{1}{128} \left(
28 - 44\cos^2\lambda + 5\sin^22\gamma\cos^4\lambda \right),
\\
j_4\left(\lambda,\gamma\right)
&=& \frac{1}{32} \left( 2 - 7\sin^22\gamma\cos^2\lambda \right)\sin2\lambda,
\\
j_5\left(\lambda,\gamma\right)
&=& \frac{1}{32} \left( 3 - 7\cos4\gamma - 7\sin^22\gamma\cos^2\lambda 
\right)\cos^2\lambda,
\\
j_6\left(\lambda,\gamma\right)
&=& \frac{1}{96} \left( 2\cos4\gamma + \sin^22\gamma\cos^2\lambda 
\right)\sin2\lambda,
\\
j_7\left(\lambda,\gamma\right)
&=& \frac{1}{1024} \left(
4\cos4\gamma\sin^2\lambda - \sin^22\gamma\cos^4\lambda \right),
\\
j_8\left(\lambda,\gamma\right)
&=& \frac{1}{32}\sin4\gamma \cos^3\lambda,
\\
j_9\left(\lambda,\gamma\right)
&=& \frac{1}{32}\sin4\gamma\cos^2\lambda\sin\lambda,
\\
j_{10}\left(\lambda,\gamma\right)
&=& \frac{1}{192} \sin4\gamma \left(5-3\cos2\lambda\right)\cos\lambda,
\\
j_{11}\left(\lambda,\gamma\right)
&=& \frac{1}{1024} \sin4\gamma\left(3-\cos2\lambda\right) \sin\lambda,
\\
j_{12}\left(\lambda,\gamma\right)
&=& \frac{1}{32} \left( 14 - \sin^22\gamma\cos^2\lambda \right) \sin2\lambda,
\\
j_{13}\left(\lambda,\gamma\right)
&=& \frac{1}{32} \left( 9 - 5\cos4\gamma - 5\sin^22\gamma\cos^2\lambda \right) 
\cos^2\lambda.
\een

\section{Monte Carlo analysis of the two signal-to-noise ratio components}

We have also studied the distribution of the individual signal-to-noise ratios
$d_1^2$ and $d_2^2$ given by Eqs.\ (\ref{snr61}) and (\ref{snr62}) over the
angles $\alpha$, $\delta$, $\psi$, and $\iota$ by means of the Monte Carlo
simulations (for the observation time $T_o = 120$ days).  The results are
shown in Figures 9--13.  The signal-to-noise ratios $d_1^2$ and $d_2^2$ are 
normalized here by means of the quantities
$$
\langle d_{1({\rm 120days})}^2\rangle
:=\langle d_1^2\rangle_{\alpha,\delta,\psi,\iota}\vert_{T_o=120{\rm days}},
\quad
\langle d_{2({\rm 120days})}^2\rangle
:=\langle d_2^2\rangle_{\alpha,\delta,\psi,\iota}\vert_{T_o=120{\rm days}}.
$$

In Tables 3 and 4 we summarize the statistical characteristics of the simulated
distributions of the normalized signal-to-noise ratios $d_1^2$ and $d_2^2$.  We
give extremal values, means, standard deviations (std), and medians.

\vspace{2ex}\begin{table}[ht]
\begin{center}
\begin{tabular}{|c|c|c|c|c|c|}\hline
detector & min & max & mean & std & median \\ \hline
GEO600          & 0.0 & 2.0 & 1.0 & 0.45 & 0.96 \\ \hline
LIGO Hanford    & 0.0 & 1.7 & 1.0 & 0.41 & 1.0 \\ \hline
LIGO Livingston & 0.0 & 1.5 & 1.0 & 0.34 & 1.1 \\ \hline
VIRGO           & 0.0 & 1.6 & 1.0 & 0.36 & 1.1 \\ \hline
TAMA300         & 0.0 & 1.9 & 1.0 & 0.38 & 1.0 \\ \hline
\end{tabular}
\end{center}
\caption{Statistics of $d_1^2/\langle d_{1({\rm 120days})}^2\rangle$.}
\end{table}

\vspace{2ex}\begin{table}[ht]
\begin{center}
\begin{tabular}{|c|c|c|c|c|c|}\hline
detector & min & max & mean & std & median \\ \hline
GEO600          & 0.18 & 4.0 & 1.0 & 0.72 & 0.77 \\ \hline
LIGO Hanford    & 0.12 & 3.3 & 1.0 & 0.68 & 0.79 \\ \hline
LIGO Livingston & 0.27 & 2.7 & 1.0 & 0.64 & 0.80 \\ \hline
VIRGO           & 0.26 & 3.2 & 1.0 & 0.66 & 0.79 \\ \hline
TAMA300         & 0.18 & 2.8 & 1.0 & 0.64 & 0.80 \\ \hline
\end{tabular}
\end{center}
\caption{Statistics of $d_2^2/\langle d_{2({\rm 120days})}^2\rangle$.}
\end{table}


\unitlength 1cm
\begin{center}\begin{figure}[ht]
\begin{picture}(10,15)
\put(0,0){\includegraphics{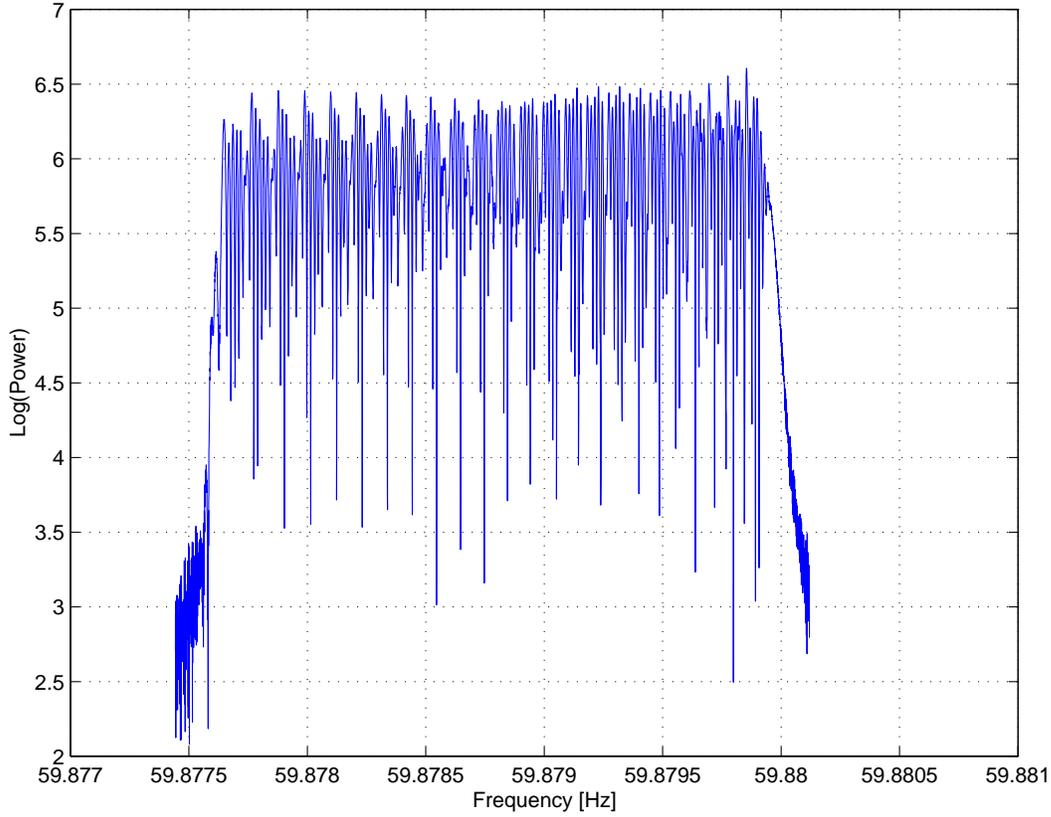}}
\end{picture}
\caption{Power spectrum of the noise-free response of an interferometer located
near Hannover to gravitational-wave signal from the Crab pulsar at twice the
rotation frequency.  We have assumed the frequency $f_0 = 29.937$ Hz and the 
spindown parameters
${\ssbii 1} = -3.773 \times 10^{-10}$ s$^{-2}$,
${\ssbii 2} = 0.976 \times 10^{-20}$ s$^{-3}$,
${\ssbii 3} = -0.615\times 10^{-30}$ s$^{-4}$. 
A 24-day long signal was analysed corresponding to the frequency resolution of 
around $4.8 \times 10^{-7}$ Hz.  The power spectrum shows 24 main peaks 
resulting from the periodic phase modulation of the signal.  In each interval 
between the main peaks there are additional subsidiary peaks arising from the 
amplitude modulation of the signal.}
\end{figure}\end{center}

\begin{center}\begin{figure}[ht]
\begin{picture}(10,13.5)
\put(0,0){\includegraphics{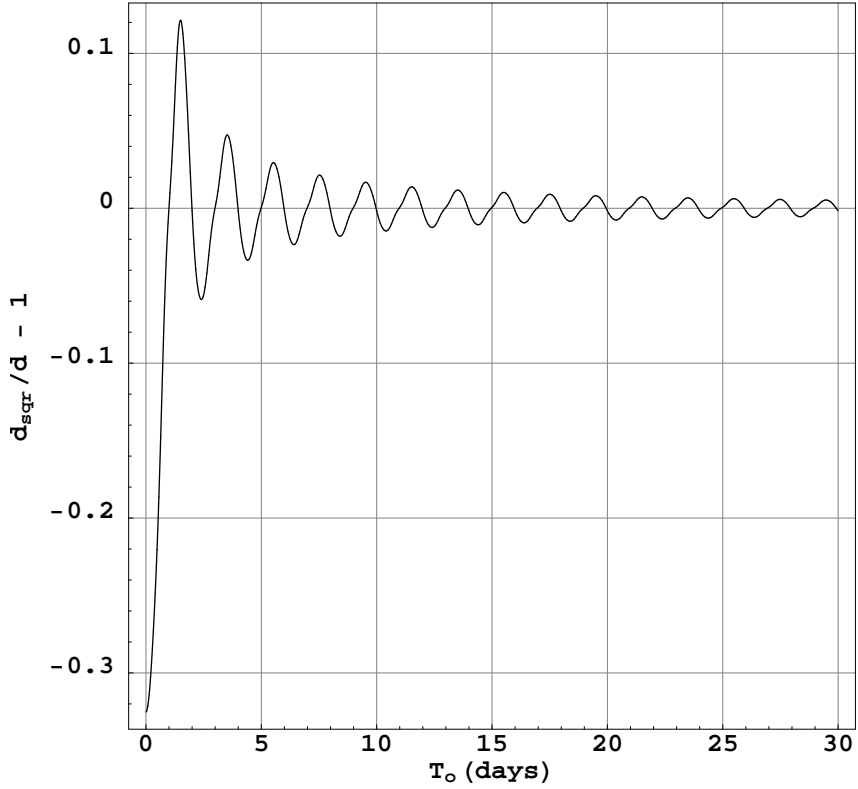}}
\end{picture}
\caption{The relative contribution of the part $d_{\rm sqr}$ of the
signal-to-noise ratio proportional to the square root of the observation time
$T_o$ to the total signal-to-noise ratio $d$ for the GEO600 detector with the
wideband noise curve ($d_{\rm sqr}\approx h_o\sin\zeta[A_1\sin^2 
2\theta/S_h(f_0) + A_2\sin^4\theta/S_h(2f_0)]^{1/2}\sqrt{T_o}$, cf.\ Eqs.\ 
(\ref{snr3}), (\ref{snr61}), and (\ref{snr62})).  A hypothetical neutron star is 
assumed to be in the distance of 40 pc from the Earth and to emit gravitational 
waves with frequency $f_0$ = 100 Hz, star's ellipticity is $10^{-5}$, its moment 
of inertia w.r.t.\ the rotation axis is $10^{45}$ g cm$^2$.  We also set
$\alpha-\phi_r=15^\circ$, $\delta=35^\circ$, $\psi=11.25^\circ$,
$\iota=22.5^\circ$, and $\theta=45^\circ$.}
\end{figure}\end{center}

\begin{center}\begin{figure}[ht]
\begin{picture}(10,10)
\put(0,0){\includegraphics{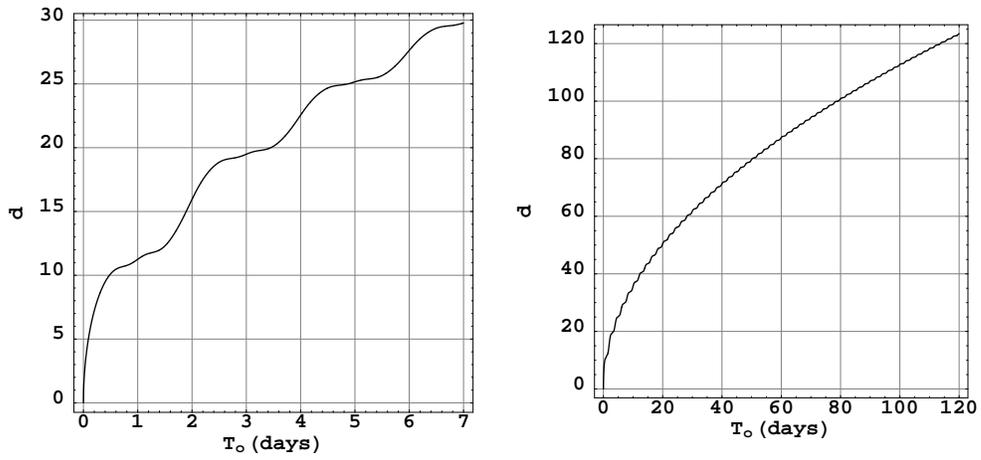}}
\end{picture}
\caption{The total signal-to-noise ratio $d$ as a function of the observation 
time $T_o$ for the GEO600 detector (with the wideband noise curve). The neutron 
star parameters are the same as in Figure 2.}
\end{figure}\end{center}

\begin{center}\begin{figure}[ht]
\begin{picture}(10,13.5)
\put(0,0){\includegraphics{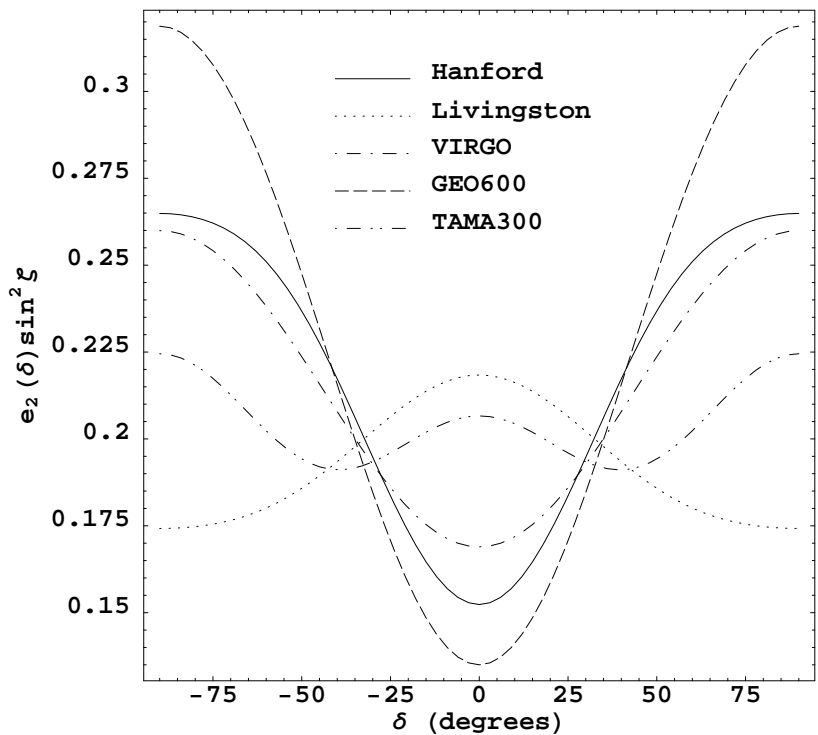}}
\end{picture}
\caption{The plot of the product $e_2(\delta)\sin^2\zeta$ against the 
declination $\delta$ of the gravitational-wave source for different detectors.
It can be shown (cf.\ Appendix B) that for a hypothetical detector located at 
the latitude $\lambda=\pm\arccos\sqrt{2/3}\approx\pm35.26^\circ$ there exist 
eight different orientations $\gamma$ of its arms such that the function 
$e_2=1/5$, i.e.\ $e_2$ does not depend on the declination $\delta$ of the 
gravitational-wave source. These orientation angles are: $\gamma_o$, 
$90^\circ\pm\gamma_o$, $180^\circ\pm\gamma_o$, $270^\circ\pm\gamma_o$, 
$360^\circ-\gamma_o$, where 
$\gamma_o=\frac{1}{2}\arcsin\sqrt{3/5}\approx25.38^\circ$.}
\end{figure}\end{center}

\begin{center}\begin{figure}[ht]
\begin{picture}(10,18)
\put(0,0){\includegraphics{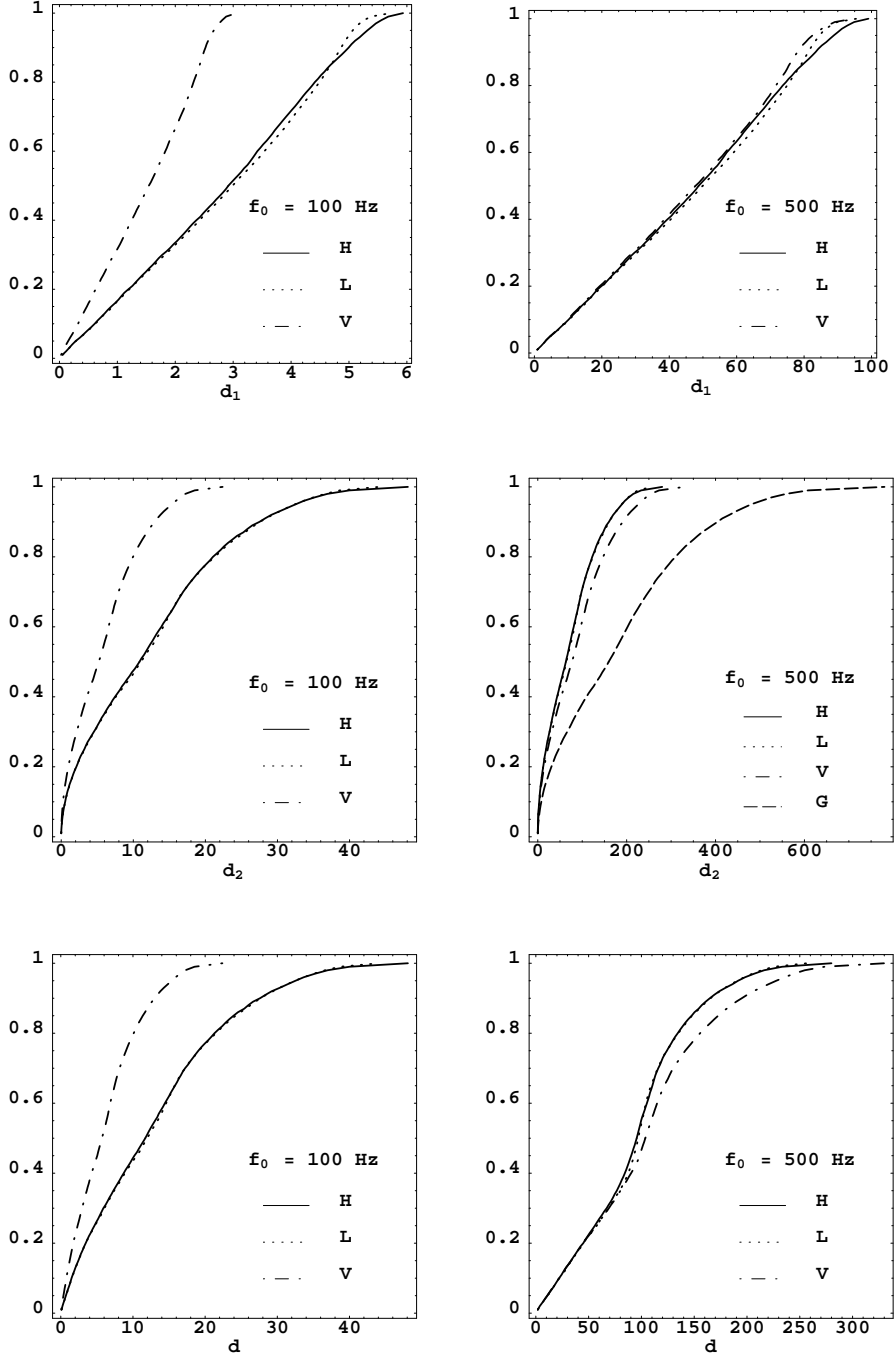}}
\end{picture}
\caption{Cumulative distribution functions of the simulated signal-to-noise 
ratios $d_1$, $d_2$, and $d$  for the VIRGO (V), initial Hanford (H), and 
initial Livingston (L) detectors.  We assume that star's ellipticity is 
$10^{-5}$, its moment of inertia w.r.t.\ the rotation axis is $10^{45}$ g 
cm$^2$ and its distance from the Earth is 1 kpc.  The observation time is 120 
days. The left column is for $f_0$ = 100 Hz and the right one is for $f_0$ = 500 
Hz. We have also shown the cumulative distribution function of the 
signal-to-noise ratio $d_2$ for the GEO600 (G) detector with the narrowband 
noise tuned to 1 kHz with the bandwidth of 30 Hz.}
\end{figure}\end{center}

\begin{center}\begin{figure}[ht]
\begin{picture}(10,18)
\put(0,0){\includegraphics{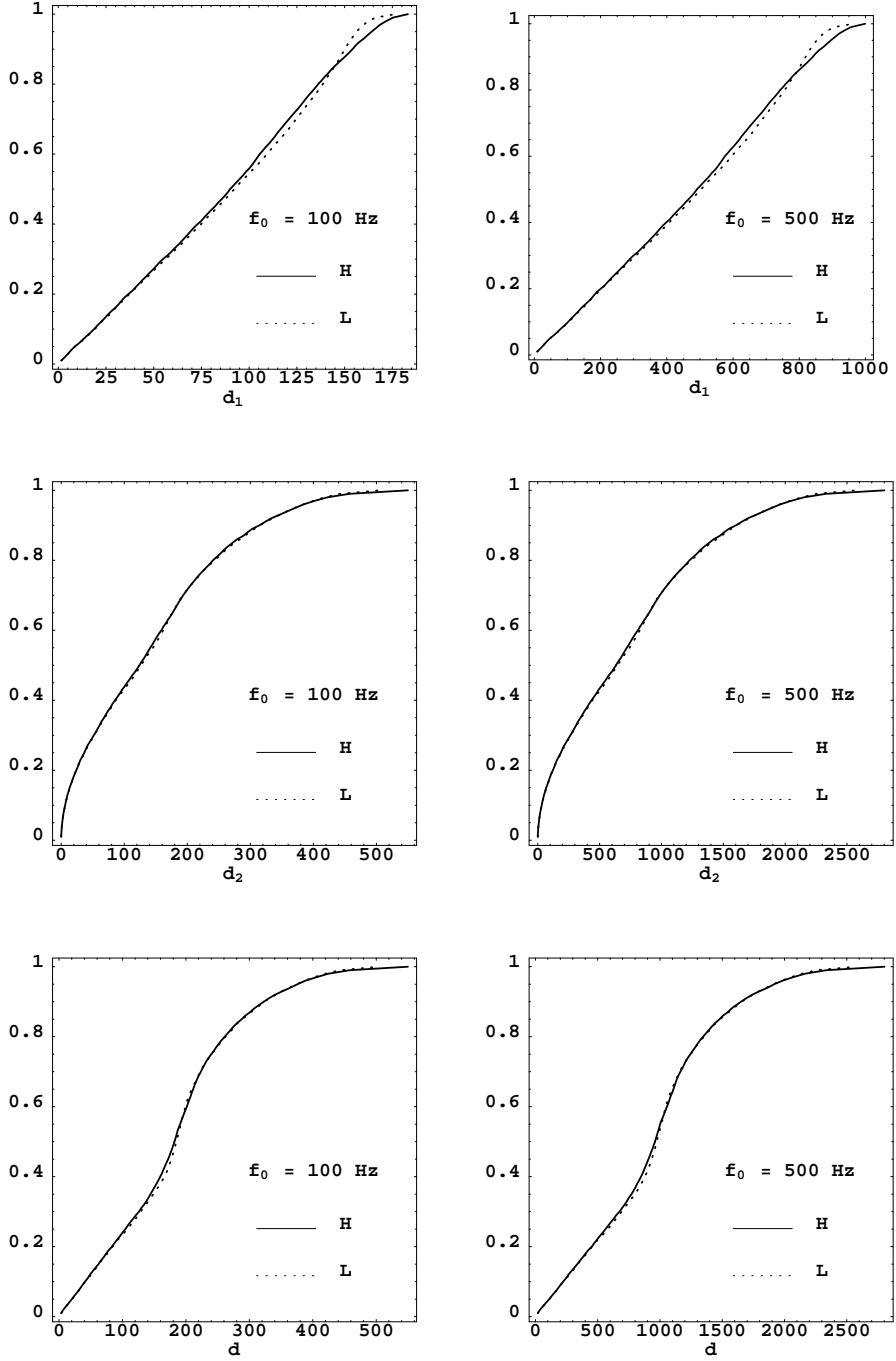}}
\end{picture}
\caption{Cumulative distribution functions of the simulated signal-to-noise 
ratios $d_1$, $d_2$, and $d$ for the advanced Hanford (H) and advanced 
Livingston (L) detectors. The observation time is 120 days. The left column is 
for $f_0$ = 100 Hz and the right one is for $f_0$ = 500 Hz. The neutron star 
parameters are the same as in Figure 5.}
\end{figure}\end{center}

\begin{center}\begin{figure}[ht]
\begin{picture}(10,18)
\put(0,0){\includegraphics{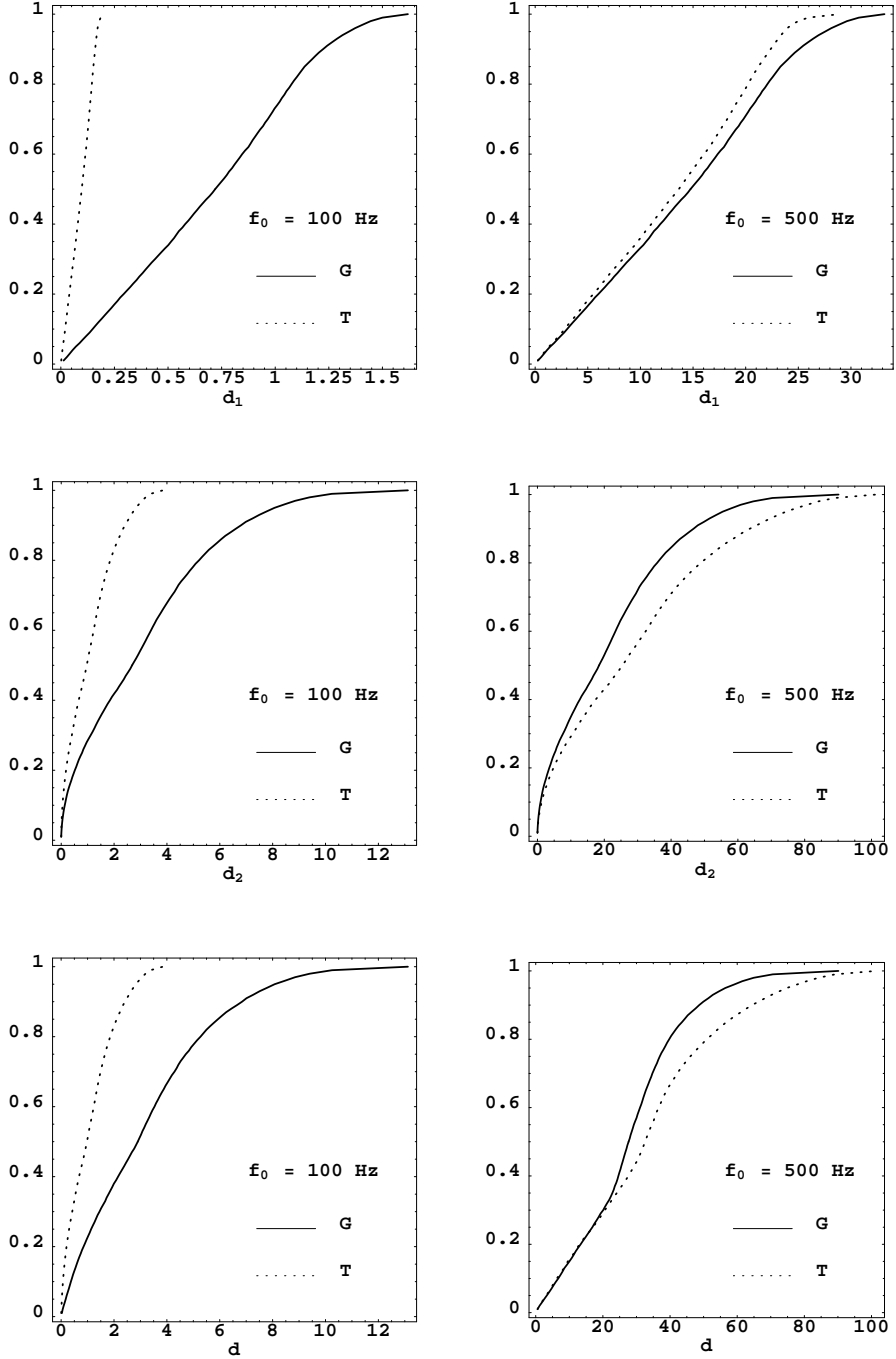}}
\end{picture}
\caption{Cumulative distribution functions of the simulated signal-to-noise 
ratios $d_1$, $d_2$, and $d$ for the GEO600 (G) with the wideband noise curve 
and TAMA300 (T) detectors. The observation time is 120 days. The left column is 
for $f_0$ = 100 Hz and the right one is for $f_0$ = 500 Hz. The neutron star 
parameters are the same as in Figure 5.}
\end{figure}\end{center}

\begin{center}\begin{figure}[ht]
\begin{picture}(10,18)
\put(0,0){\includegraphics{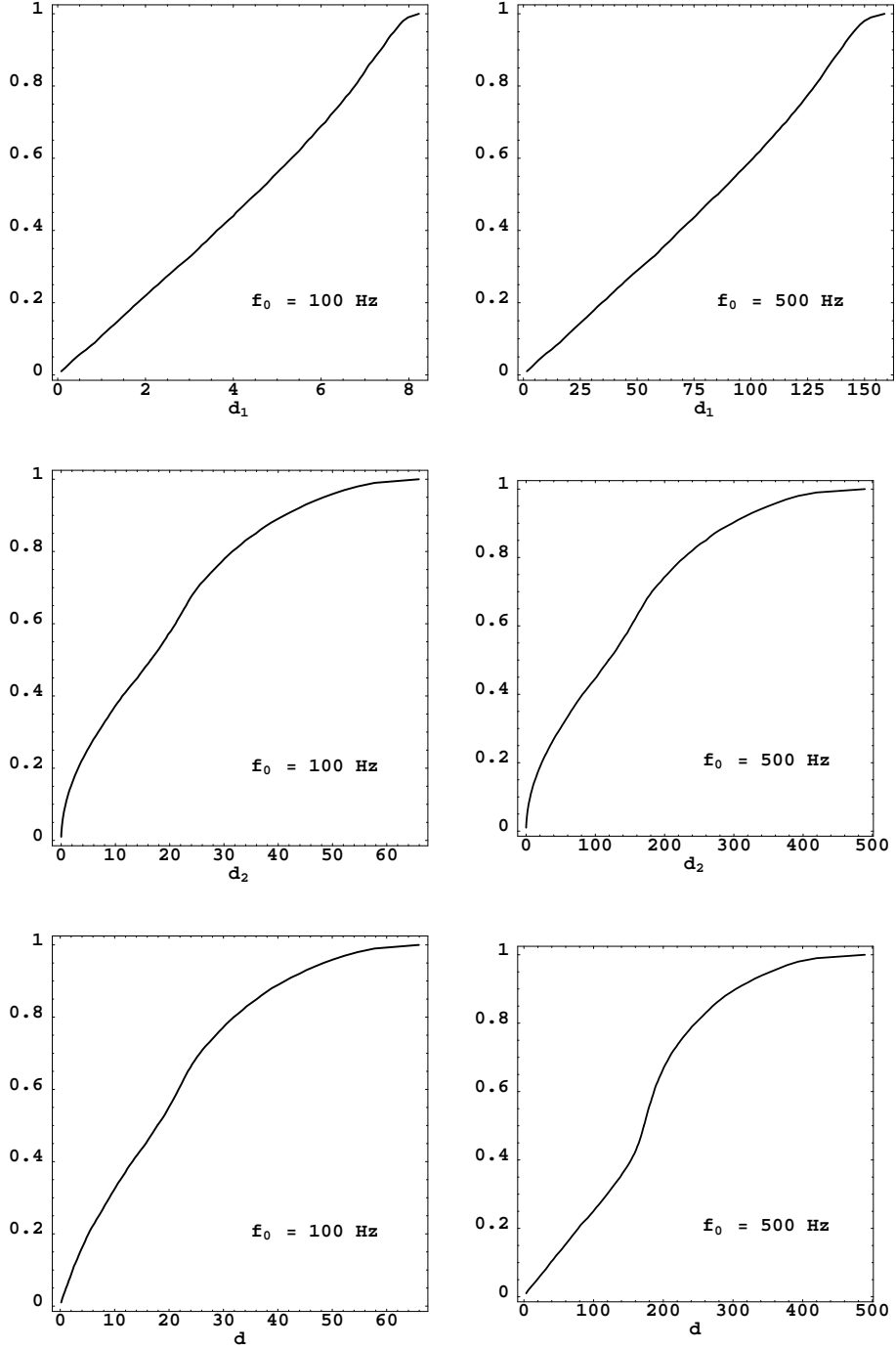}}
\end{picture}
\caption{Cumulative distribution functions of the simulated signal-to-noise 
ratios $d_{n1}$, $d_{n2}$, and $d_n$ for the three detector network of the VIRGO 
and two initial LIGO detectors. The observation time is 120 days. The left 
column is for $f_0$ = 100 Hz and the right one is for $f_0$ = 500 Hz. The 
neutron star parameters are the same as in Figure 5.}
\end{figure}\end{center}

\begin{center}\begin{figure}[ht]
\begin{picture}(10,15)
\put(0,0){\includegraphics{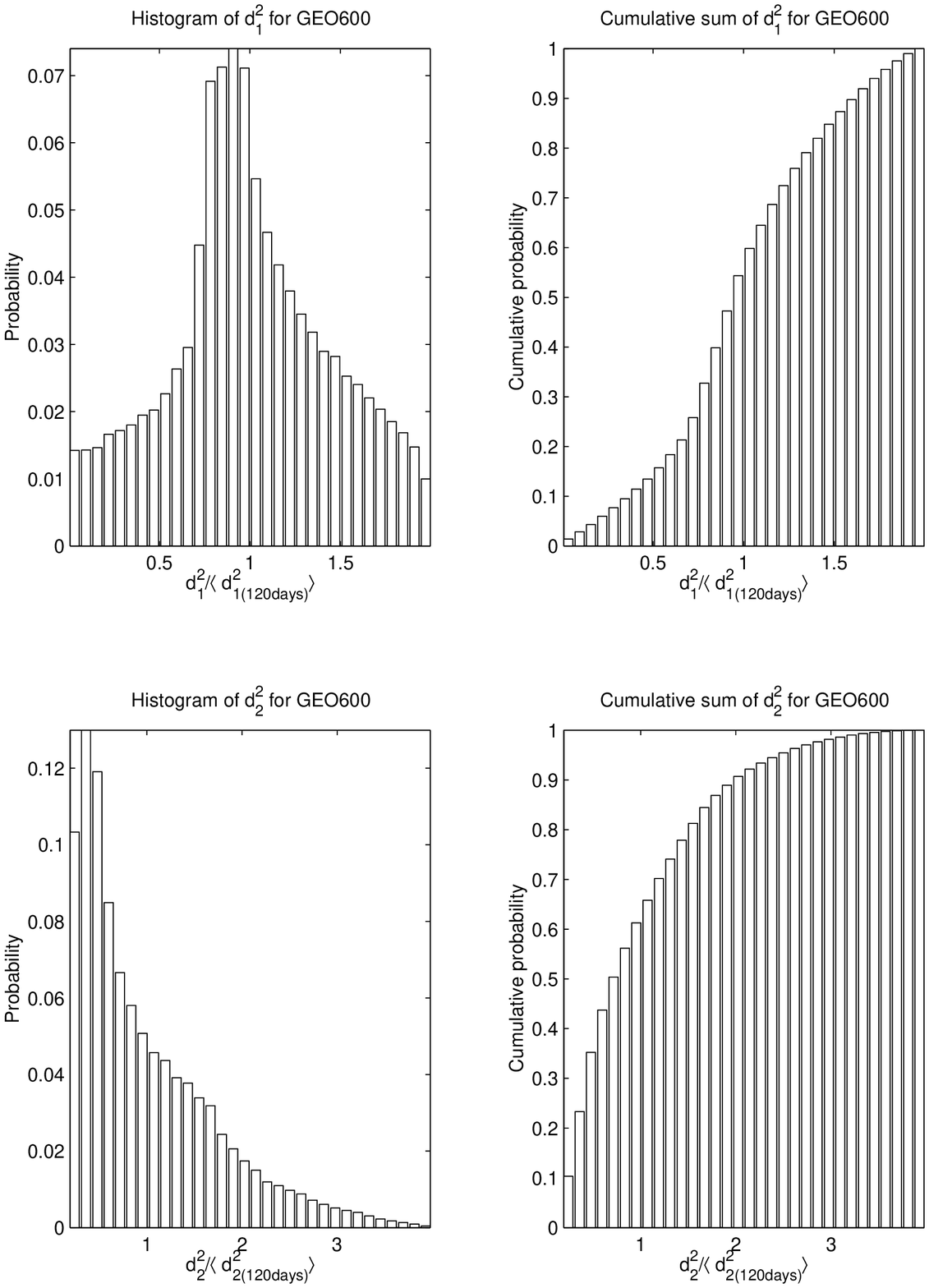}}
\end{picture}
\caption{Histograms of the simulated probability density and cumulative 
distribution functions of the normalized signal-to-noise ratios 
$d_1^2/\langle d_{1({\rm 120days})}^2\rangle$ and
$d_2^2/\langle d_{2({\rm 120days})}^2\rangle$ for the GEO600 detector.}
\end{figure}\end{center}

\begin{center}\begin{figure}[ht]
\begin{picture}(10,15)
\put(0,0){\includegraphics{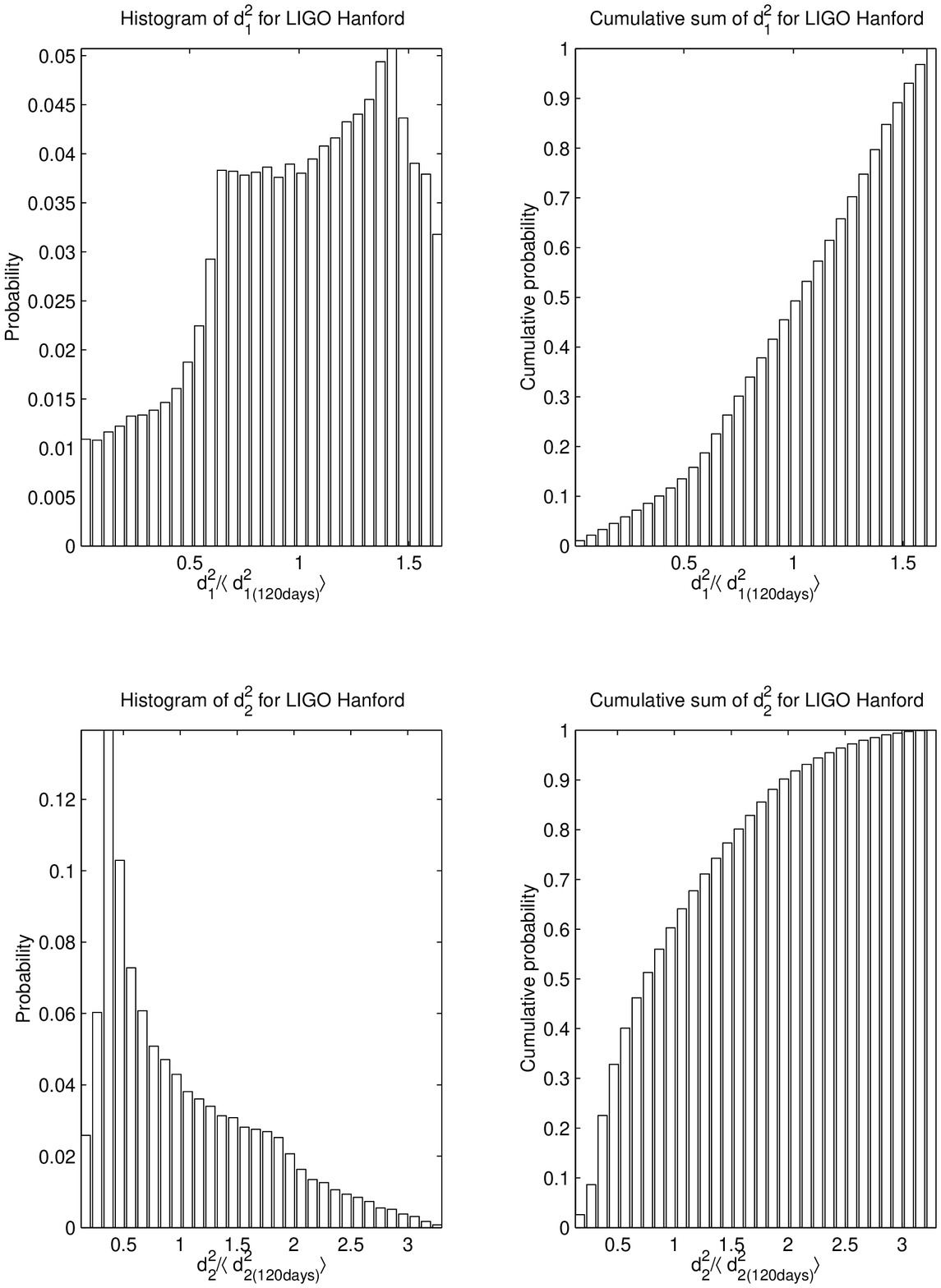}}
\end{picture}
\caption{Histograms of the simulated probability density and cumulative 
distribution functions of the normalized signal-to-noise ratios 
$d_1^2/\langle d_{1({\rm 120days})}^2\rangle$ and
$d_2^2/\langle d_{2({\rm 120days})}^2\rangle$ for the LIGO Hanford detector.}
\end{figure}\end{center}

\begin{center}\begin{figure}[ht]
\begin{picture}(10,15)
\put(0,0){\includegraphics{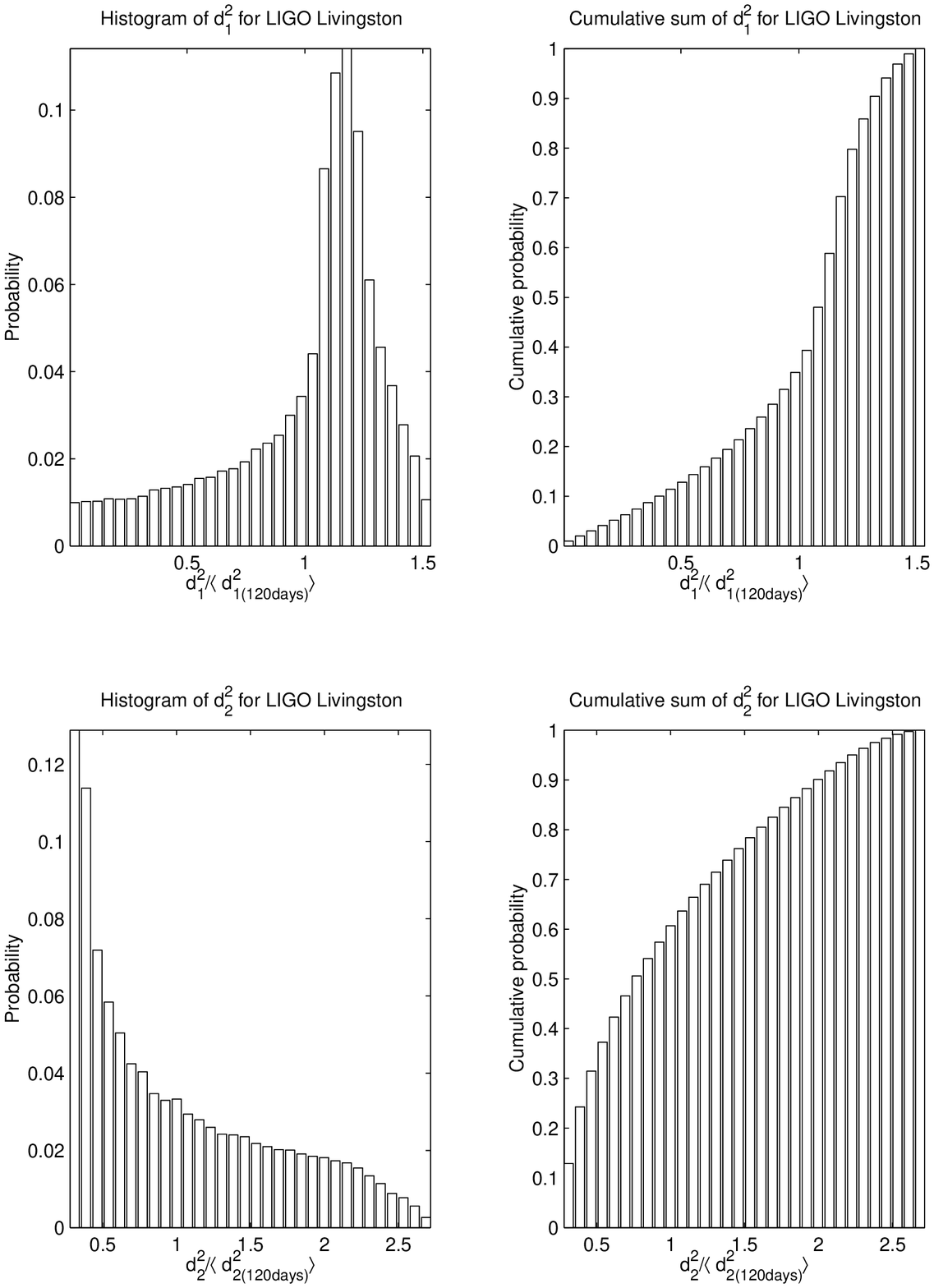}}
\end{picture}
\caption{Histograms of the simulated probability density and cumulative 
distribution functions of the normalized signal-to-noise ratios 
$d_1^2/\langle d_{1({\rm 120days})}^2\rangle$ and
$d_2^2/\langle d_{2({\rm 120days})}^2\rangle$ for the LIGO Livingston detector.}
\end{figure}\end{center}

\begin{center}\begin{figure}[ht]
\begin{picture}(10,15)
\put(0,0){\includegraphics{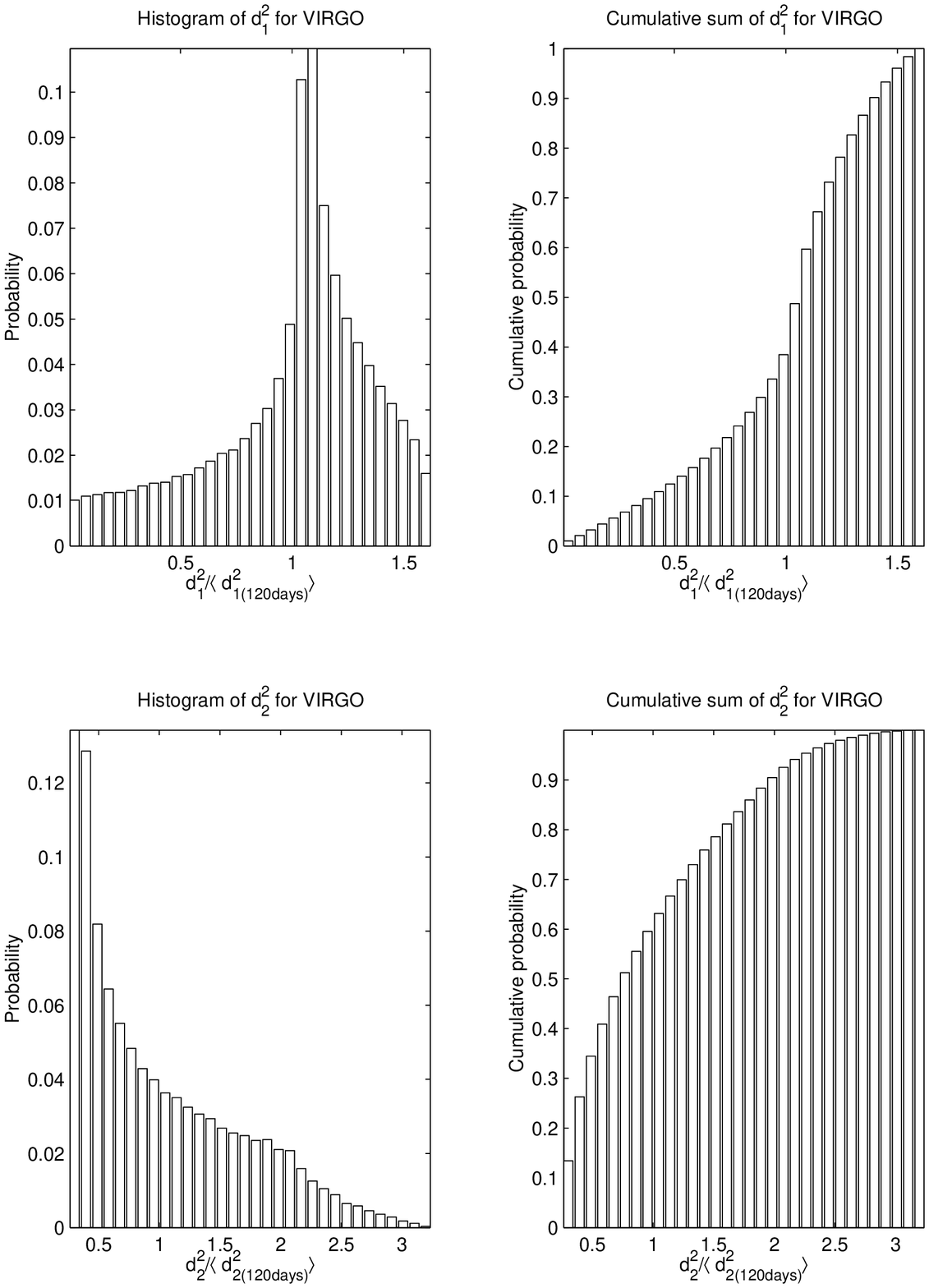}}
\end{picture}
\caption{Histograms of the simulated probability density and cumulative 
distribution functions of the normalized signal-to-noise ratios  
$d_1^2/\langle d_{1({\rm 120days})}^2\rangle$ and
$d_2^2/\langle d_{2({\rm 120days})}^2\rangle$ for the VIRGO detector.}
\end{figure}\end{center}

\begin{center}\begin{figure}[ht]
\begin{picture}(10,15)
\put(0,0){\includegraphics{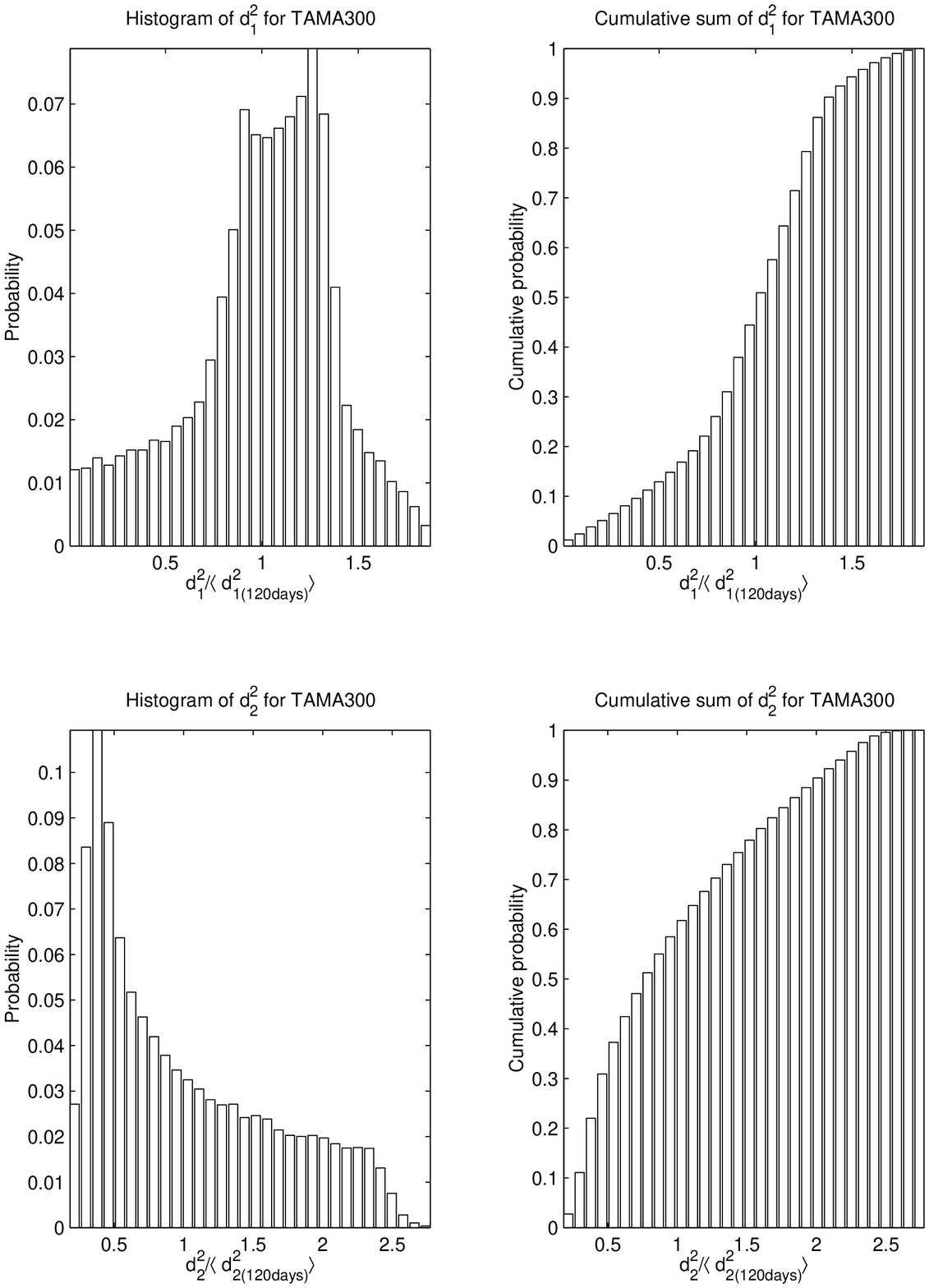}}
\end{picture}
\caption{Histograms of the simulated probability density and cumulative 
distribution functions of the normalized signal-to-noise ratios 
$d_1^2/\langle d_{1({\rm 120days})}^2\rangle$ and
$d_2^2/\langle d_{2({\rm 120days})}^2\rangle$ for the TAMA300 detector.}
\end{figure}\end{center}


\begin{thebibliography}{99}

\bibitem{S96a}
B.\ F.\ Schutz,
in {\em Proceedings of the September 1996 SIGRAV Meeting}
(World Scientific, Singapore, 1997).

\bibitem{T87}
K.\ S.\ Thorne,
in {\em Three Hundred Years of Gravitation},
edited by S.\ W.\ Hawking and W.\ Israel
(Cambridge University Press, Cambridge, 1987),
Section 9.4.2(b),
pp.\ 386--391.

\bibitem{GEO600}
K.\ Danzmann {\it et al.},
in {\it Gravitational Wave Experiments},
edited by E.\ Coccia, G.\ Pizzella, and F. Ronga
(World Scientific, Singapore, 1995),
pp.\ 100--111.

\bibitem{LIGO}
A.\ Abramovici {\it et al.},
Science {\bf 256}, 325 (1992).

\bibitem{VIRGO}
C.\ Bradaschia {\it et al.},
Nucl.\ Instrum.\ Methods Phys.\ Res.\ A {\bf 289}, 518 (1990).

\bibitem{TAMA300}
K.\ Tsubono {\it et al.},  
in {\it Gravitational Wave Detection. Proceedings of the TAMA International 
Workshop on Gravitational Wave Detection},
edited by K.\ Tsubono, M.-K.\ Fujimoto, and K.\ Kuroda
(Universal Academy Press, Tokyo, 1997),
pp.\ 183--191.

\bibitem{PPS76}
V.\ R.\ Pandharipande, D.\ Pines, and R.\ A.\ Smith,
Astrophys.\ J.\ {\bf 208}, 550 (1976).

\bibitem{BG96}
S.\ Bonazzola and E.\ Gourgoulhon,
Astron.\ Astr.\ {\bf 312}, 675 (1996).

\bibitem{ZS79}
M.\ Zimmermann and E.\ Szedenits,
Phys.\ Rev.\ D {\bf 20}, 351 (1979).

\bibitem{Z80}
M.\ Zimmermann,
Phys.\ Rev.\ D {\bf 21}, 891 (1980).

\bibitem{W84}
R.\ V.\ Wagoner,
Astrophys.\ J.\ {\bf 278}, 345 (1984).

\bibitem{S96b}
B.\ F.\ Schutz,
in {\em Mathematics of Gravitation. Part II: Gravitational Wave Detection},
edited by A.\ Kr\'olak
(Banach Center Publications Vol.\ 41 Part II, Warsaw, 1997),
pp.\ 11--17.

\bibitem{C70}
S.\ Chandrasekhar,
Phys.\ Rev.\ Lett.\ {\bf 24}, 611 (1970).

\bibitem{FS78}
J.\ L.\ Friedman and B.\ F.\ Schutz,
Astrophys.\ J.\ {\bf 222}, 281 (1978).

\bibitem{S91}
B.\ F.\ Schutz,
in {\em The Detection of Gravitational Waves},
edited by D.\ G.\ Blair
(Cambridge University Press, Cambridge, 1991),
pp.\ 406--452.

\bibitem{And97}
N.\ Andersson, A new class of unstable modes of rotating relativistic
stars, preprint gr-qc/9706075, 1997.

\bibitem{Lin97}
L.\ Lindblom,
in {\em Relativistic Astrophysics},
edited by H. Riffert {\em et al.}
(Vieweg Verlag, Wiesbaden, 1997).

\bibitem{L87}
J.\ C.\ Livas, Ph.\ D.\ thesis,
Massachusetts Institute of Technology, 1987.

\bibitem{N93}
T.\ M.\ Niebauer {\em et al.},
Phys. Rev. D {\bf 47}, 3106 (1993).

\bibitem{J95}
G.\ S.\ Jones, Ph.\ D.\ thesis,
University of Wales, 1995.

\bibitem{BCCS97}
P.\ R.\ Brady, T.\ Creighton, C.\ Cutler, and B.\ F.\ Schutz,
Phys. Rev. D {\bf 57}, 2101 (1998).

\bibitem{K97a}
A.\ Kr\'olak,
in {\it Very High Energy Phenomena in the Universe.
Proceedings of the XXXIInd Rencontres de Moriond},
edited by Y.\ Giraud-H\'eraud and J.\ Tr\^an Thanh V\^an
(Editions Frontieres, Paris, 1997),
pp.\ 329--334.

\bibitem{ST87}
B.\ F.\ Schutz and M.\ Tinto,
Mon.\ Not.\ R.\ Astron.\ Soc.\ {\bf 224}, 131 (1987).

\bibitem{JK94}
P.\ Jaranowski and A.\ Kr\'olak,
Phys.\ Rev.\ D {\bf 48}, 1723 (1994).

\bibitem{A96}
B.\ Allen,
Gravitational wave detector sites,
gr-qc 9607075.

\bibitem{JD94}
K.\ Jotania and S.\ V.\ Dhurandhar,
Bull.\ Astr.\ Soc.\ India {\bf 22}, 303 (1994).

\bibitem{N93a}
Our functions $a$ and $b$ from Eqs.\ (\ref{adef}) and (\ref{bdef}) are connected 
with the functions $S_+$ and $S_\times$ from Eqs.\ (9) and (10) of \cite{N93} by 
relations: $S_+=-4a$, $S_\times=4b$, provided the following identification of 
our variables $\alpha,\delta,\lambda,\phi_r,\gamma$ with the variables 
$\alpha,\beta,\theta,\psi,\lambda$ used in \cite{N93} is made (assuming 
$\zeta=\pi/2$): $\alpha\to\beta$, $\delta\to\pi/2-\alpha$, $\phi_r\to\lambda$, 
$\lambda\to\pi/2-\theta$, $\gamma\to\psi-5\pi/4$. The last identification means 
that the angle $\psi$ is measured clockwise from the first inteferometer's arm 
to North.

\bibitem{Liv}
J.\ C.\ Livas,
in {\em Gravitational Wave Data Analysis},
edited by B.\ F.\ Schutz (Kluwer, Dordrecht, 1989),
pp.\ 217--238.

\bibitem{O96}
B. J.\ Owen,
Phys.\ Rev.\ D {\bf 53}, 6749 (1996).

\bibitem{DSch}
S.\ V.\ Dhurandhar and B.\ F.\ Schutz,
Phys.\ Rev.\ D {\bf 50}, 2390 (1994).

\bibitem{Da}
M.\ H.\ A.\ Davis,
in {\it Gravitational Wave Data Analysis},
edited by B.\ F.\ Schutz (Kluwer, Dordrecht, 1989),
pp.\ 73--94.

\bibitem{H68}
C.\ W.\ Helstr\"om,
{\it Statistical Theory of Signal Detection}, 2nd ed.
(Pergamon Press, London, 1968),
Sections 2(c), 2(d), and 3 of Chapter IX.

\bibitem{VN}
D.\ Nicholson and A.\ Vecchio,
Bayesian Bounds on Parameter Estimation Accuracy 
for Compact Coalescing Binary Gravitational Wave Signal,
preprint gr-qc/9705064, 1997. 

\bibitem{Cal}
C.\ Cutler {\em et al.},
Phys.\ Rev.\ Lett.\ {\bf 70}, 2984 (1993).

\bibitem{S96}
B.\ F.\ Schutz,
Class.\ Quantum Grav.\ {\bf 13}, A219 (1996).

\bibitem{SKpc}
S.\ Kawamura (private communication).

\bibitem{KSpc}
K.\ Strain (private communication).

\bibitem{BD}
R. Balasubramanian and S. V.  Dhurandhar,
Phys. Rev. D {\bf 50}, 6080 (1994).

\bibitem{S}
B. S. Sathyaprakash,
Phys. Rev. D {\bf 50}, R7111 (1994). 

\bibitem{Kc}
A. Kr\'olak,
in {\it Proceedings of the Cornelius Lanczos International
Centenary Conference},
edited by J. D. Brown, M. T. Chu, D. C. Ellison, and R. J. Plemmons
(SIAM, Philadelphia, 1994), p. 482. 

\bibitem{BH86}
D.\ C.\ Backer and R.\ W.\ Hellings,
Ann.\ Rev.\ Astron.\ Astrophys.\ {\bf 24}, 537 (1986).

\bibitem{CS67}
G.\ M.\ Clemence and V.\ Szebehely,
Astron.\ J.\ {\bf 72}, 1324 (1967).

\bibitem{BT76}
R.\ Blandford and S.\ A.\ Teukolsky,
Astrophys.\ J.\ {\bf 205}, 580 (1976).

\bibitem{S64}
I.\ I.\ Shapiro,
Phys.\ Rev.\ Lett.\ {\bf 13}, 789 (1964).

\end{thebibliography}
\end{document}